\documentclass[namedreferences]{solarphysics}
\usepackage[optionalrh]{spr-sola-addons}
\usepackage{amssymb}
\usepackage{graphicx}
\usepackage{epstopdf}
\newcommand\ion[2]{#1$\;${\scshape{#2}}}
\usepackage[position=top,justification=raggedright,singlelinecheck=false]{subfig}
\usepackage{textcomp}

\begin{document}
\begin{article}
\begin{opening}

\title{Quiescent and Eruptive Prominences at Solar Minimum: A Statistical Study \textit{via} an Automated Tracking System}

\author{I.P.~\surname{Loboda}$^{1}$\sep S.A.~\surname{Bogachev}$^{1}$ }
\institute{$^1$ P.N.~Lebedev Physical Institute of the Russian Academy of Sciences, Moscow, Russia \\ email: \url{loboda@sci.lebedev.ru}}
\runningauthor{I.P. Loboda and S.A. Bogachev}
\runningtitle{Quiescent and Eruptive Prominences at Solar Minimum}

\begin{abstract}

We employ an automated detection algorithm to perform a global study of solar prominence characteristics.
We process four months of TESIS observations in the \ion{He}{ii} 304~\AA\ line taken close to the solar minimum of 2008--2009 and focus mainly on quiescent and quiescent-eruptive prominences.
We detect a total of 389 individual features ranging from $25 \times 25$ to $150 \times 500$~Mm$^2$ in size and obtain distributions of many their spatial characteristics, such as latitudinal position, height, size and shape.
To study their dynamics, we classify prominences as either stable or eruptive and calculate their average centroid velocities, which are found to be rarely exceeding 3~km~s$^{-1}$.
Besides, we give rough estimates of mass and gravitational energy for every detected prominence and use these values to evaluate the total mass and gravitational energy of all simultaneously existing prominences ($10^{12}$--$10^{14}$~kg and $10^{29}$--$10^{31}$~erg, respectively).
Finally, we investigate the form of the gravitational energy spectrum of prominences and derive it to be a power-law of index~$-1.1 \pm 0.2$.

\end{abstract}

\end{opening}

\section{Introduction}

Prominences are one of the most noticeable features of the Sun, which, although observed for over a century, are still far from being completely understood.
For historical reasons, a distinction is made between prominences, observed off-limb as luminous formations, and filaments, usually seen in absorption on the disk.
Physically, these are the same structures consisting of plasma with properties similar to those of the chromosphere being nearly 100 times denser and cooler than the surrounding corona \citep{1985SoPh..100..415H, 1995tandberg, 2010SSRv..151..243L}.
Prominences vary largely in their size, morphology, dynamics, magnetic structure and the way they are related to the rest of the solar activity.
As more of these differences were revealed, new classification schemes were introduced by a number of authors \citep{1932ApJ....76....9P, 1953severny, 1966soat.book.....Z, 1984A&A...131...33L}.
In this paper, we will follow the most general one, which divides prominences into two major classes depending on whether they are part of an active region (active prominences), or whether they exist separately over the quiet Sun (quiescent ones).
It is known that instabilities in the magnetic structure supporting both these types of prominences can result in eruption of some part of their material high into the corona \citep{1989ApJ...343..971V, 2013ApJ...777...52Z}.
To such prominences we will refer here as eruptive ones.

Comparative studies have shown that more than 70\% of prominence eruptions are followed by coronal mass ejections, or CMEs \citep{2000ApJ...537..503G, 2003ApJ...586..562G, 2008AnGeo..26.3025F}.
They, in turn, travel over vast distances from the Sun and have a considerable impact on the heliosphere in general, including the Earth's magnetosphere.
At the same time, the exact connection between prominence eruptions and CMEs, as well as the mechanisms of formation and eruption of prominences themselves, are still poorly understood.
Existing models of prominence formation consider a number of possible magnetic topologies and mechanisms that can fill them with plasma, while observational data are not sufficient to discriminate between these models \citep{1999ApJ...510..444G, 2010SSRv..151..333M}.
Similarly, theories of prominence eruptions developed so far suffer from poorly constrained pre-eruptive conditions of the magnetic field and plasma \citep{2003NewAR..47...53L}. 
In this context, statistical studies of prominences seem particularly useful as they may provide the necessary conditions and constraints to theories and establish reliable correlations of prominence evolution with other manifestations of solar activity.

For decades, prominences have been observed from the Earth in a number spectral lines, primarily H$_{\alpha}$, as well as in radio emission \citep{2001A&A...380..323L, 2007ASPC..370...46H}.
A comprehensive study of these phenomena, however, became only possible with the advent of space-based observations.
Nowadays, prominences can be observed from space by a number of instruments and in a much wider spectral range.
Above all, these observations are free of day-night interruptions and unfavourable atmospheric effects.
Most of them are performed in the EUV spectral range, where prominences are best seen in the \ion{He}{ii} 304~\AA\ line.
Despite the vast amounts of data produced by these instruments, most of the present-day studies of prominences limit themselves to only examining individual, and usually, the largest events.
In this way, however, one risks missing smaller prominences, which are of great interest still, and obtaining an incomplete picture of these phenomena.
Therefore, to make use of all the data available, and to study the entire set of prominences, special detection and tracking algorithms need to be applied. 

A continuous set of observations produced by the \textit{Extreme ultraviolet Imaging Telescope} (EIT) of the \textit{Solar and Heliospheric Observatory} (SOHO), the \textit{Extreme Ultraviolet Imager} (EUVI) of the \textit{Solar TErrestrial RElations Observatory} (STEREO) and the \textit{Atmospheric Imaging Assembly} (AIA) of the \textit{Solar Dynamics Observatory} (SDO) since 1996 now covers a vast period of almost two solar cycles.
Nevertheless, only a few attempts have been made so far to use these data for large statistical studies of prominences.
Two algorithms have been developed for EIT data \citep{2006SoPh..234..135F, 2010SoPh..262..449L} and another one for STEREO-A and -B observations \citep{2010ApJ...717..973W}.
Furthermore, only the last attempt produced a  prominence catalogue that is now publicly available on-line.
Continuing this line of study, we have developed a new algorithm capable of precisely locating prominences using monochromatic observations in the \ion{He}{ii} 304~\AA\ line only.
We employ this algorithm to study a period of four months of observations, which corresponds roughly to the end of the prolonged solar minimum of 2008--2009. 
Hence, we focus mainly on quiescent and quiescent-eruptive prominences and leave active prominences for future studies.

The organization of this paper is as follows: in Section~\ref{sect:observations} we present the observational data used and in Section~\ref{sect:processing} we give a general description of the detection algorithm employed.
We then review the major results of its operation and perform statistical analysis of the obtained prominence parameters in Section~\ref{sect:results}.
Finally, we discuss the system's performance, as well as advantages and future prospects of automated detection of prominences in Section \ref{sect:discussion}.

\section{Observations}\label{sect:observations}

For this study, we use observations from one of the two \textit{Full-disk EUV Telescopes} (FET) of the TESIS solar observatory, namely, the FET-304/171 telescope  \citep{2009AdSpR..43.1001K}.
This instrument was designed at the Lebedev Institute of the Russian Academy of Sciences and launched on board the \textit{Complex ORbital Observations Near-Earth of Activity of the Sun-Photon} (CORONAS-Photon) satellite on 30 January 2009.
To our advantage, CORONAS-Photon's relatively short operation time (until mid January 2010) coincided with the end of an extraordinarily long solar minimum between Solar Cycles 23 and 24.
This gave us an opportunity to unrestrictively observe small prominences, barely seen during the rest of the cycle against larger manifestations of solar activity.
Here, we analyse the most complete portion of this dataset from August to November 2009, which is publicly available on-line at \href{http://tesis.lebedev.ru/en/}{\texttt{http://tesis.lebedev.ru}}.

\begin{figure}[t]
	\centering
	\includegraphics[width=\linewidth]{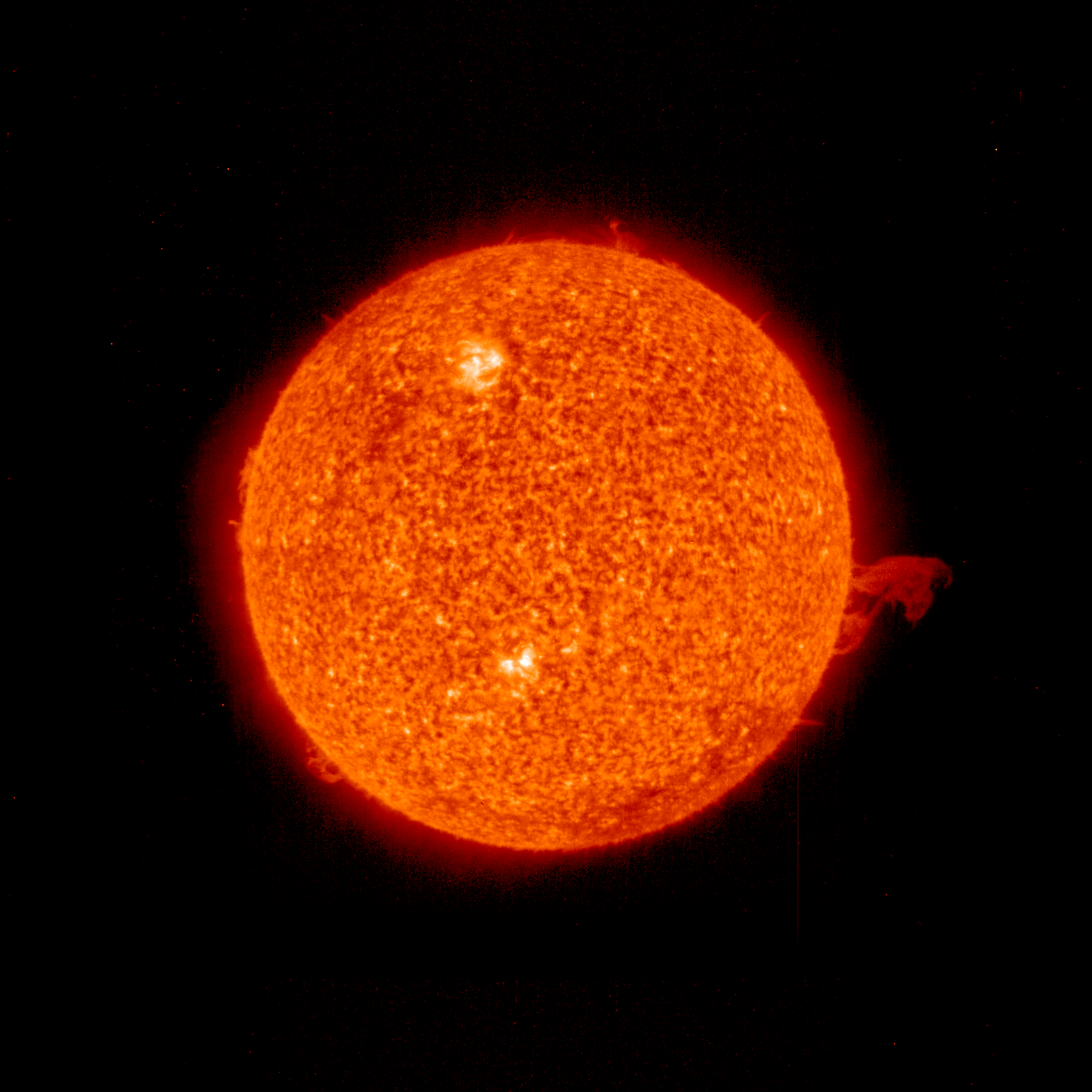} \vspace{-3mm}
	\caption{Example of typical TESIS observation in \ion{He}{ii} 304~\AA\ line taken on 26 September 2009 at 15:44:46~UT (intensities are in logarithmic scale). Note that solar North is not up in the original images.}
	\label{fig:full_disk}
\end{figure}

\begin{figure}[t]
	\centering \vspace{-4mm}
	\subfloat[\vspace{-2mm}]{\label{fig:CCD_resp}\includegraphics[]{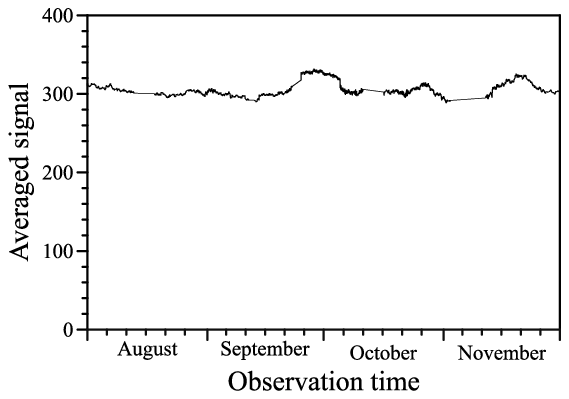}}
	\subfloat[\vspace{-2mm}]{\label{fig:304_refl}\includegraphics[]{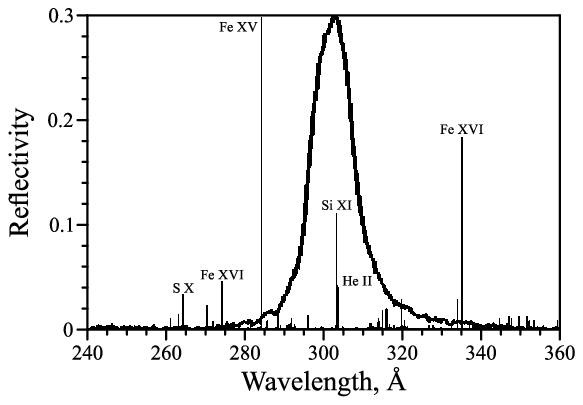}}
	\caption{\protect\subref{fig:CCD_resp} Averaged signal from the CCD throughout the observation period; we have made the correction for varying Sun-Earth distance. \protect\subref{fig:304_refl} Spectral reflectivity of the focusing mirror in the 304~\AA\ channel of TESIS and the corresponding EUV lines.}
\end{figure}

With a field of view of $1^{\circ}$, the telescope was able to image the full solar disk and the corona up to the distance of $\sim0.95$ solar radii above the limb (see an example of a typical observation in Figure~\ref{fig:full_disk}).
It produced 2048$\times$2048 images with a pixel size of $\sim1.7$~arcsec, or roughly 1.25~Mm.
Half of the images, however, have been reduced to 1024$\times$1024 due to telemetry limitations.
For the regular observations in the 304~\AA\ channel the cadence was set to 4.0 min, but as the observations were non-continuous because of the partly Earth-shaded orbit of the satellite, the resulting mean cadence amounted to $\sim7.2$ min.
To make sure that the CCD showed no apparent signs of degradation during the observation period, which could affect our investigations, we calculate the average response of the CCD, shown in Figure~\ref{fig:CCD_resp}.
It clearly shows that possible degradation of the CCD is negligible against the larger variations in solar irradiation.

Since the entrance and detector filters of the telescope were relatively broad, dedicated mainly to block the strong white-light radiation from the Sun, the bandpass in the 304~\AA\ channel of TESIS was predominantly defined by the spectral reflectivity of its focusing mirror, which is shown in Figure~\ref{fig:304_refl} \citep{2011SoSyR..45..162K}.
This figure also indicates the main contributors to this relatively broad-band channel: the closely spaced \ion{Si}{xi} line at 303.3~\AA\ and \ion{He}{ii} line at 303.8~\AA.
Contribution of the \ion{Si}{xi} line has been previously investigated by several authors, who have found it to be around 3--4\% for the on-disk quiet Sun regions and more than 90\% for the coronal emission \citep{1978SoPh...58..299C,2000SoPh..195...45T}.
Since plasma conditions in prominences are similar to that of the quiet Sun, helium emission is typically dominant in them.

\section{Image Processing}\label{sect:processing}

A  total of 18812 valid images of the Sun in the \ion{He}{ii}~304~\AA\ line were obtained during the specified period.
As we mentioned above, processing such an amount of data, including detection of prominences and measurement of their parameters, is not feasible by hand.
For this purpose, we have developed an automatic method, capable of locating the prominences on individual 304~\AA\ images, tracking these features throughout their lifespan, and determining their most important spatial and dynamic characteristics.
The operational principle of this method is described below as a sequence of several processing steps.

\subsection{Preprocessing and Background Removal}

\begin{figure}[t]
	\centering \vspace{-1mm}
	\subfloat[\vspace{-5mm}]{\hspace{5mm}\label{fig:steps_polar}\includegraphics[width=\dimexpr\linewidth-5mm]{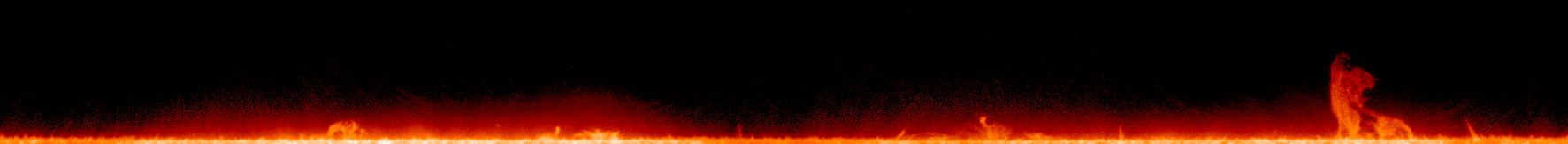}}		
	\\ \vspace{-2mm}
	\subfloat[\vspace{-5mm}]{\hspace{5mm}\label{fig:steps_bgrem}\includegraphics[width=\dimexpr\linewidth-5mm]{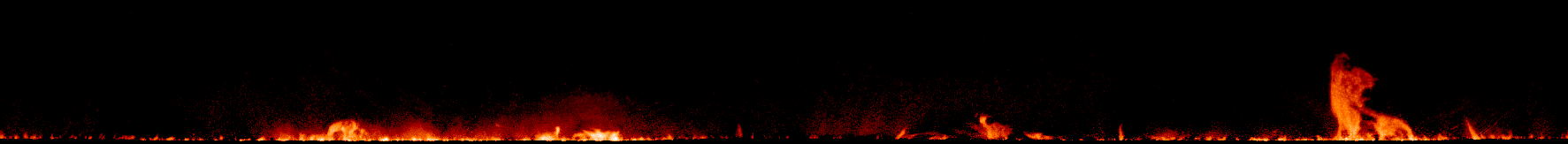}}
	\\ \vspace{-2mm}
	\subfloat[\vspace{-5mm}]{\hspace{5mm}\label{fig:steps_norm}\includegraphics[width=\dimexpr\linewidth-5mm]{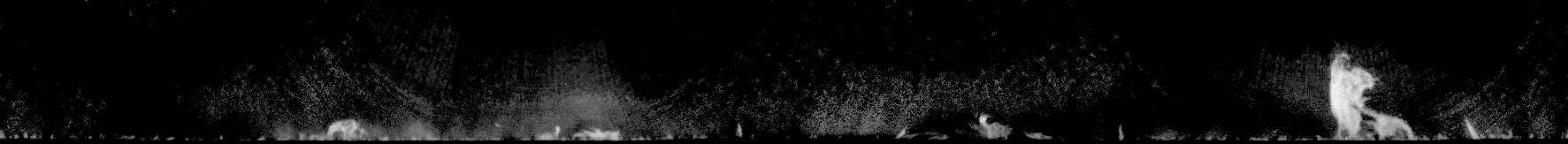}}
	\\ \vspace{-2mm} 
	\subfloat[\vspace{-5mm}]{\hspace{5mm}\label{fig:steps_kern}\includegraphics[width=\dimexpr\linewidth-5mm]{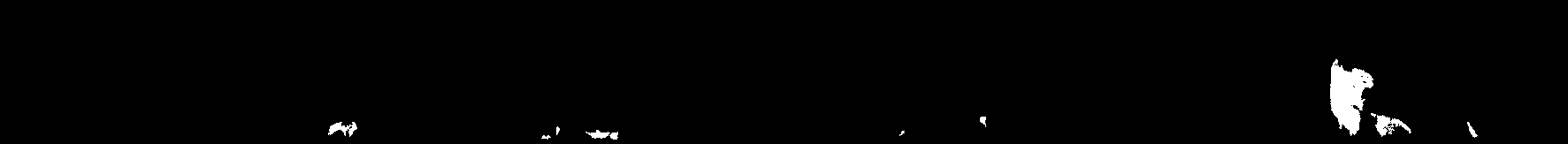}}
	\\ \vspace{-2mm}
	\subfloat[\vspace{-5mm}]{\hspace{5mm}\label{fig:steps_mask}\includegraphics[width=\dimexpr\linewidth-5mm]{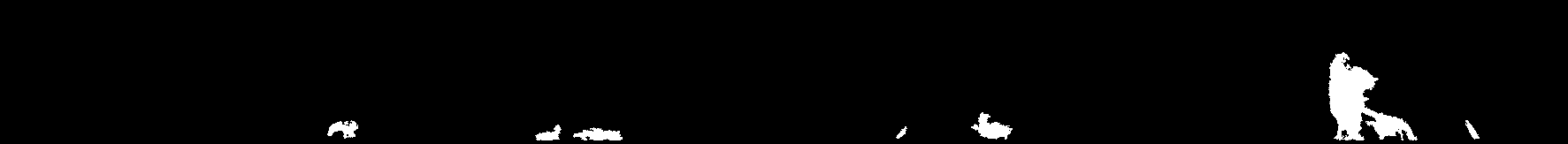}}
	\vspace{1mm}
	\caption{Processing steps of the image in Figure~\ref{fig:full_disk}: \protect\subref{fig:steps_polar} polar-transformed image with minor defects corrected, latitudes counted clockwise from the North pole, \protect\subref{fig:steps_bgrem} background subtracted, \protect\subref{fig:steps_norm} normalized image, \protect\subref{fig:steps_kern} kernels used for the subsequent region-growing procedures, \protect\subref{fig:steps_mask} final selections. Intensities in panels a to c are in logarithmic scale.}
\end{figure}

In the preliminary stage, the data is prepared for subsequent operations: corrupted images are filtered out and for the remainder corrections for minor defects are made (\textit{e.g.}, hot pixels and cosmic ray hits).
Apart from that, off-limb regions of the images are transformed to polar coordinates and stored as rectangular arrays in order to simplify further image processing (for an example, see Figure~\ref{fig:steps_polar}).

\begin{figure}[t]
	\centering \vspace{-4mm}
	\subfloat[\vspace{-2mm}]{\label{fig:rad_prof}\includegraphics{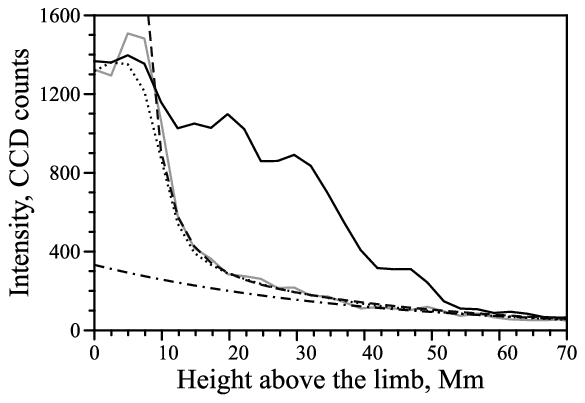}}
	\subfloat[\vspace{-2mm}]{\label{fig:int_stats}\includegraphics{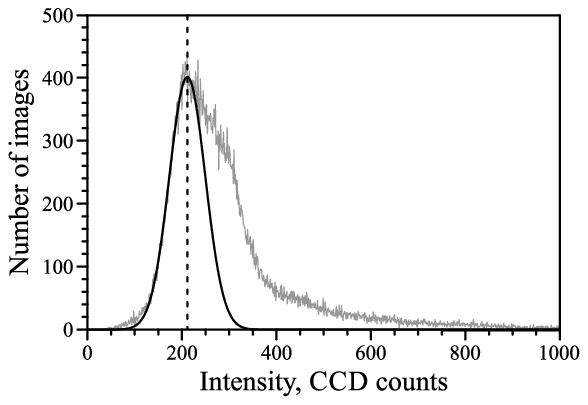}}
	\caption{ \protect\subref{fig:rad_prof} Radial profiles of the quiet Sun (solid grey line), prominence (solid black line), statistically obtained background (dotted line), analytical fit to that background (dashed line) and its exponential part (dashed-dotted line).
\protect\subref{fig:int_stats} Intensity distribution at a fixed point of corona (grey line) with an assumed Gaussian contribution from the quiet Sun (black line).}
\end{figure}

Essentially, the next step would be to detect prominences on these pre-processed images.
Unfortunately, as one can see from the Figure~\ref{fig:rad_prof}, prominences are observed against a strong background, which can be attributed either to the thermal noise of the CCD and extra signal due to the scattered radiation, or to the radiation of the undisturbed corona itself.
As \cite{2010SoPh..262..449L} rightly note, to know the background is critical to the whole process. 
Analysis of individual images shows, that this background does not change much with time, but is spatially highly inhomogeneous: it rapidly decreases with height and varies with latitude, having significant dips around the poles.
To our advantage again, because the observation period is close to the solar minimum, the positions of the polar coronal holes remain stable in spite of the solar rotation. 
This allows us to determine the background following the statistical approach described below.

First, we use the whole dataset to build the intensity distribution of every individual pixel of the previously polar-transformed and aligned images, an example of which is shown in Figure~\ref{fig:int_stats}. 
We assume that in the quiet Sun, the observed intensity coming from a fixed point in the corona follows a normal distribution.
This quiet Sun distribution is supplemented with stronger signals coming from a small number of active features, which occasionally emerge in the corona. %in the domain of larger intensities 
Therefore, we approximate the left slope of the obtained distribution with the Gaussian function and take the position of its maximum as the estimated value of background in that pixel.

In order to smooth out the noise and small-scale irregularities, we then look for an analytical function to fit the background we have thus obtained.
Having studied its radial and latitudinal profiles on semi-log and log-log plots, we have finally arrived to the following model:
\begin{equation}
B(\varphi, h) = a(\varphi)e^{ -b(\varphi)h + c(\varphi)e^{-d(\varphi)h} } \,,
\label{eq:bgmodel}
\end{equation}
where $h$ is the height above the limb, $\varphi$ is the heliographic latitude and $a(\varphi)$, $b(\varphi)$, $c(\varphi)$ and $d(\varphi)$ are generalized logistic functions of a similar form, \textit{e.g.}
\begin{equation}
a(\varphi) = a_1 + \frac{a_2 - a_1}{e^{ \left(\varphi_0 - \left| \varphi \right| \right)/\delta\varphi }+1} \,.
\end{equation}
In other words, $a(\varphi)$ is equal to $a_1$ at the equator and middle latitudes and to $a_2$ in the vicinity of poles, with a smooth transition from $a_1$ to $a_2$ centred at the latitude $\varphi_0$, the steepness of this transition determined by $\delta\varphi$. 
Parameters $\varphi_0$ and $\delta\varphi$ are taken the same for all four functions.

Although the double exponent in Equation~(\ref{eq:bgmodel}) may seem strange at a first glance, it proved to be the best approximation to the specific radial profiles of the background obtained for this dataset; while $B(h)$ has an almost exponential dependence at higher altitudes, it becomes much steeper close to the limb (Figure~\ref{fig:rad_prof}).
Moreover, below a height of $\sim15$~Mm the background is mostly due to the strong radiation from the EUV spicules and prominences are in general not visible there.
Therefore, we do not attempt to fit this part with an analytical function.

Subsequently, we follow a technique similar to that described in \cite{2006SoPh..236..263M} and use this modelled background $B$ to normalize every image in the dataset according to the equation:
\begin{equation}
N= \frac{I+\delta}{B+\delta} = 1 + \frac{I-B}{B+\delta} \,,
\end{equation}
where $I=I(\varphi, h)$ is a given polar-transformed image and $N=N(\varphi, h)$ is the same image after normalization (see an example in Figure~\ref{fig:steps_norm}).
We introduce here a small parameter $\delta$, first, to avoid division by zero, and second, to suppress the noise at higher altitudes, where $B$ is close to zero.
Thus, it must be substantially greater than the average noise level in the periphery of the image ($\sim5$ CCD counts for this instrument) and substantially smaller than typical signal values from prominences ($\sim100$ and more CCD counts). 
For the dataset employed, we have set $\delta$ equal to~25.

\subsection{Prominence Detection in the Individual Images}\label{sect:detection}

To avoid restrictions imposed by the necessity of simultaneous multi-wave\-length observations, our first concern was to be able to detect prominences using 304~\AA\ observations only.
The problem is, however, that they are not the only bright structures seen off-limb in this usually broad-band channel.
High coronal temperatures above the active regions (ARs) can stimulate emission in the \ion{Si}{xi} line and cause a considerable increase in brightness above the limb.
Consequently, additional procedures are required to discriminate between true prominences and these AR-imposed effects.
A number of solutions to this problem have been previously proposed by several authors.
\cite{2006SoPh..234..135F}, for example, implemented a histogram segmentation to set an intensity threshold for an individual image and then employed multi-wavelength observations to discard active regions.
\cite{2010ApJ...717..973W} first selected potential prominences by a preset threshold and then performed a regression analysis based on the shape of the selected regions.
A disadvantage of this method is, however, that wrong thresholding may lead to misclassification and therefore to no prominence detection.
Finally, an interesting method was proposed by \cite{2010SoPh..262..449L}; these authors used moments to find radial profiles characteristic of prominences only and then reconstructed their shape by morphological opening procedures.

We have developed  a different technique, taking advantage of the fact that, in the \ion{He}{ii} 304~\AA\ line, quiescent prominences tend to have sharp boundaries (at least in the dataset concerned), and on the contrary, active regions reveal themselves off-limb as diffuse brightenings.
In our study, we concentrate mainly on the measurement of the physical parameters of prominences for the later statistical treatment.
Thus, our primary concern was to achieve high precision in prominence detection in order to derive their properties most reliably.
In order to meet this requirement, our algorithm individually determines an optimal threshold for each prominence on the image by the following procedure.

\begin{figure}[t]
	\centering \vspace{-4mm}
	\subfloat[\vspace{-2mm}]{\label{fig:grow_prom}\includegraphics[]{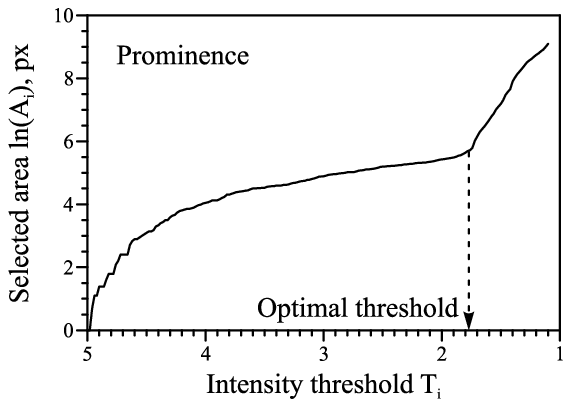}}
	\subfloat[\vspace{-2mm}]{\label{fig:grow_AR}\includegraphics[]{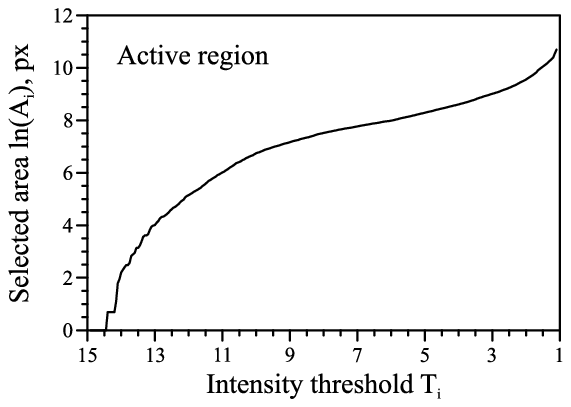}}
	\caption{Selected area $A_i$ as it grows with the decrease of the cut-off threshold $T_i$ in case of a prominence \protect\subref{fig:grow_prom} and an active region \protect\subref{fig:grow_AR}.}
\end{figure}

First, kernels of future prominences are found to isolate individual off-limb features.
These are the areas where the normalized signal $N$ is greater than a fixed threshold $T_{kern}=3.0$.
Each kernel is then treated separately and expanded by means of the region-growing procedure.
This operation is performed multiple times with a varying cut-off threshold $T_i$, which is gradually decreased from $T_{max} = 7.0$ to $T_{min} = 1.1$ at an increment of $\delta T= -0.02$.
Concurrently, the area of the resulting selection $A_i$ is measured as a function of $T_i$.
Naturally, as $T_i$ decreases, $A_i$ grows monotonically; a typical view of this dependence is shown in Figure~\ref{fig:grow_prom}.
One can notice that $T_i$ can be greater than $T_{kern}$.
In this case, the region-growing procedure is reduced to finding all  the fitted points inside the kernel.
In addition, if in the region-growing process two kernels merge together, their measured dependencies $A_i(T_i)$ are summed.

\begin{figure}[t]
	\centering \vspace{-3mm}
	\subfloat[\vspace{-1mm}$T_i=7.0$]{\label{fig:grow_70}\includegraphics[width=0.24\linewidth]{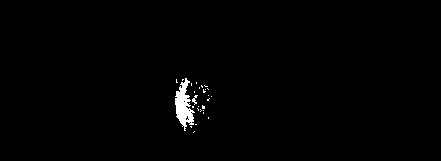}} \hfill
	\subfloat[\vspace{-1mm}$T_i=6.0$]{\label{fig:grow_60}\includegraphics[width=0.24\linewidth]{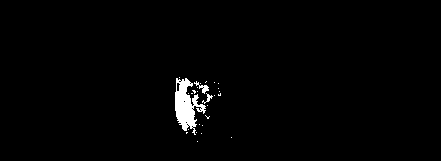}} \hfill	
	\subfloat[\vspace{-1mm}$T_i=5.0$]{\label{fig:grow_50}\includegraphics[width=0.24\linewidth]{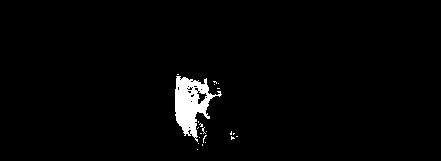}} \hfill
	\subfloat[\vspace{-1mm}$T_i=4.0$]{\label{fig:grow_40}\includegraphics[width=0.24\linewidth]{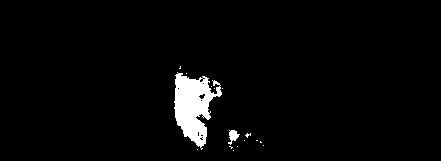}}
	\\ \vspace{-2mm}
	\subfloat[\vspace{-1mm}$T_i=3.0$]{\label{fig:grow_30}\includegraphics[width=0.24\linewidth]{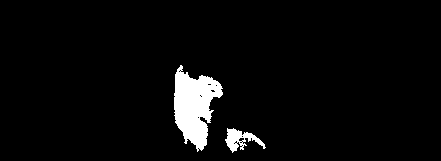}} \hfill	
	\subfloat[\vspace{-1mm}$T_i=2.0$]{\label{fig:grow_20}\includegraphics[width=0.24\linewidth]{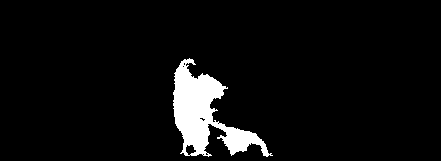}} \hfill
	\subfloat[\vspace{-1mm}$T_i=1.6$]{\label{fig:grow_16}\includegraphics[width=0.24\linewidth]{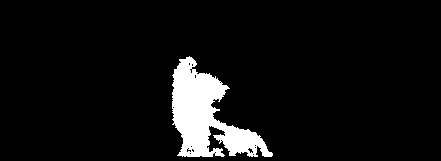}} \hfill	
	\subfloat[\vspace{-1mm}$T_i=1.5$]{\label{fig:grow_15}\includegraphics[width=0.24\linewidth]{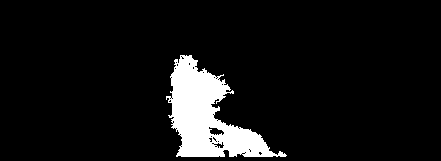}}
	\\ \vspace{-2mm}
	\subfloat[\vspace{-1mm}$T_i=1.4$]{\label{fig:grow_14}\includegraphics[width=0.24\linewidth]{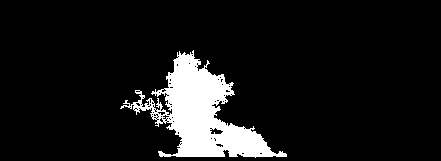}} \hfill
	\subfloat[\vspace{-1mm}$T_i=1.3$]{\label{fig:grow_13}\includegraphics[width=0.24\linewidth]{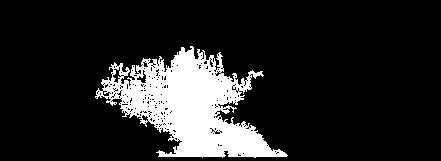}} \hfill	
	\subfloat[\vspace{-1mm}$T_i=1.2$]{\label{fig:grow_12}\includegraphics[width=0.24\linewidth]{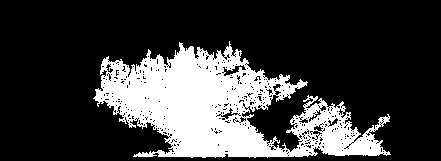}} \hfill	
	\subfloat[\vspace{-1mm}$T_i=1.1$]{\label{fig:grow_11}\includegraphics[width=0.24\linewidth]{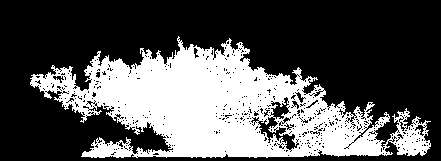}}
	\caption{Resulting selection at several stages of the iterative region-growing procedure.}
	\label{fig:grow_proc}
\end{figure}

Initially, when the main body of a prominence is being selected, $A_i$ grows quite rapidly (Figures~\ref{fig:grow_proc}\subref*{fig:grow_70}--\subref*{fig:grow_40}).
Later, when the sharp boundaries of the prominence are reached, the increase of $A_i$ slows down (Figures~\ref{fig:grow_proc}\subref*{fig:grow_30}--\subref*{fig:grow_15}); this is easily identifiable in Figure~\ref{fig:grow_prom} as some sort of a saturation plateau.
Finally, when $T_i$ becomes sufficiently low such that regions surrounding the prominence begin to be selected, $A_i$ bursts to a rapid growth (Figures~\ref{fig:grow_proc}\subref*{fig:grow_14}--\subref*{fig:grow_11}).
The value of $T_i$ preceding this burst is then chosen as an optimal intensity threshold for the given prominence.

\begin{figure}[t]
	\centering
	\includegraphics[width=0.24\linewidth]{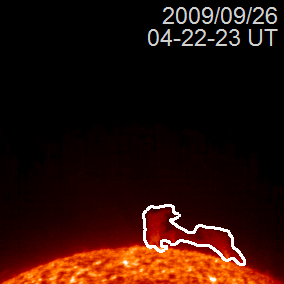} \hfill
	\includegraphics[width=0.24\linewidth]{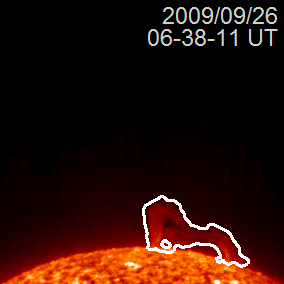} \hfill
	\includegraphics[width=0.24\linewidth]{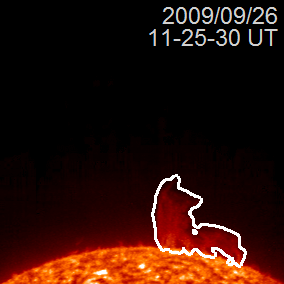} \hfill
	\includegraphics[width=0.24\linewidth]{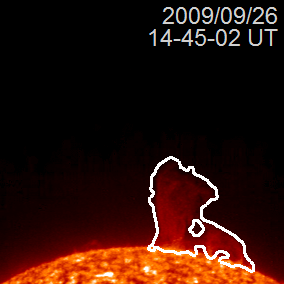}
	\\ \vspace{0.01\linewidth}
	\includegraphics[width=0.24\linewidth]{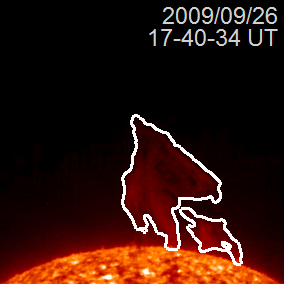} \hfill
	\includegraphics[width=0.24\linewidth]{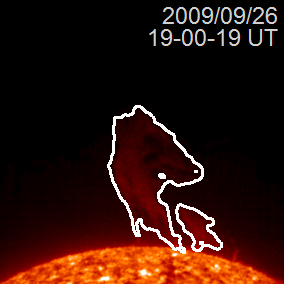} \hfill
	\includegraphics[width=0.24\linewidth]{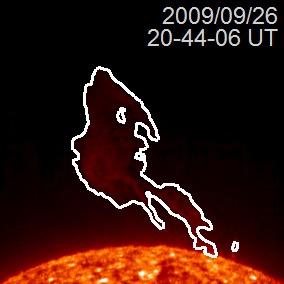} \hfill
	\includegraphics[width=0.24\linewidth]{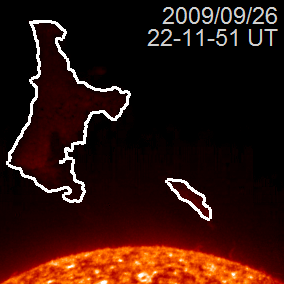}
	\caption{The high dynamic range of detection exemplified by a large eruptive event observed on north-eastern limb on 26 September 2009. Result of the detection outlined with a bright curve; intensities are in logarithmic scale.}
\label{fig:erpt}
\end{figure}

The main advantage of this approach is the high dynamic range of detection.
This is well illustrated by Figure~\ref{fig:erpt}; at the late stages of eruption the faint prominence material is detected just as well as the bright body of the prominence before the eruption has started.
Another important feature of this technique is that it allows the direct discrimination between prominences and active regions.
As the latter have no sharp boundaries, the dependence $A_i(T_i)$ is smoother, with no characteristic knee as is the case of a prominence (Figure~\ref{fig:grow_AR}).
Therefore, no optimal threshold is determined and they are not detected as prominences.
This, of course, inhibits the detection of active prominences, that are often inseparable from the active regions, but as we state above, they are not the subject of the present study.

We should note, however, that this technique is still biased by the arbitrary choice of $T_{kern}$.
Fortunately, this parameter is chosen such that it would fall into the saturation plateau in Figure~\ref{fig:grow_prom}, which precedes the burst of the selected area.
Since the plateau is broad enough, there is no difficulty to choose $T_{kern}$, which will fall into this interval for most of the prominences.
Therefore, we believe that small variations of $T_{kern}$ are not relevant for the algorithm performance.
When testing the algorithm, we have eventually chosen $T_{kern} = 3.0$ since it produced better results for a small portion of the dataset that we have studied manually.
As for the computational efficiency of the algorithm, the most time-consuming operation is the iterative region-growing procedure described above.
However, since each successive iteration does not expand the initial seed, but continues to expand the result of the previous iteration, the whole process can be interpreted as a single process, with a number of intermediate stops at certain thresholds.
Therefore, the actual processing time is much smaller than that required to preform the same number of independent region-growing procedures.
In our case, the processing of each image took typically from 2 to 3 seconds on a 3.1~GHz eight-core CPU, depending on the number of kernels initially selected.

\subsection{Parameter Extraction, Event Tracking, and Catalogue Compilation}
Once all prominences are duly located on the image, for each of them, heliographic latitude and altitude of the highest point are determined, and visible (\textit{i.e.} projected) area, integral brightness, and position of the centroid are calculated.
Furthermore, a set of models described below in Section~\ref{sect:models} is used to estimate its volume, mass and, therefore, its gravitational energy.
Additionally, in order to find the erupted parts of prominences, selections that are detached from the disk and are entirely located above a certain height of 30~Mm are marked.

Eventually, the last step of our algorithm is to track the same prominences as they appear on a series of consecutive images.
We notice that most of the prominences are confined to the same latitudes during their entire lifetime.
Therefore, two prominences can be considered as the same event if they overlap on two or more consecutive images.
They can be absent, however, on no more than two images in a row and must be observed on a minimum of three images.
Having tracked a prominence with this approach, the algorithm registers the moment of its appearance and disappearance, calculates its lifetime and the number of images in which it was detected.
For all of the characteristics of a prominence determined previously on the individual images, the minimum, the maximum, the average, and the standard deviation are found.
After that, tangential and radial velocities of the centroid of a prominence are calculated based on the change of its position, as well as the radial velocity of the highest point of the prominence.
Finally, if the tracked prominence had at any time regions that were marked as erupted, the whole event is classified as eruptive prominence.
All of these data are stored in a table, which is the final result of the application of the algorithm.
In the next section, we briefly describe these results for the observation period and perform a statistical analysis of the obtained properties of the prominences.

To evaluate the accuracy of the detection, we have manually investigated a relatively short period of ten days from 10 to 20 September 2009.
During this period, a total of 46 prominences was detected.
We found observable prominences in the images for 42 of these detections, although some of them were barely distinguishable by eye.
For the remaining four, we could find no features resembling an off-limb prominence.
Most likely, these false detections resulted from the increased coronal radiation due to the presence of an internetwork bright point close to the limb, or from a superposition of short-lived and small-scale phenomena, like surges and polar macrospicules.
At the same time, we have found no large prominences that were not detected by the algorithm; it is still possible, however, that we could have missed several rather small ones.
Besides, all 42 prominences were correctly classified by the algorithm as either stable or eruptive (although, for smaller prominences it proved to be harder to draw this distinction).
The algorithm was sometimes unable to detect faint remainders or precursors of the prominences, therefore the derived lifetimes are most likely to be slightly underestimated.

\section{Results}\label{sect:results}
\subsection{Dynamics and Morphology}

\begin{figure}[t]
	\centering \vspace{-4mm}
	\subfloat[\vspace{-2mm}]{\label{fig:distrib_len}\includegraphics[]{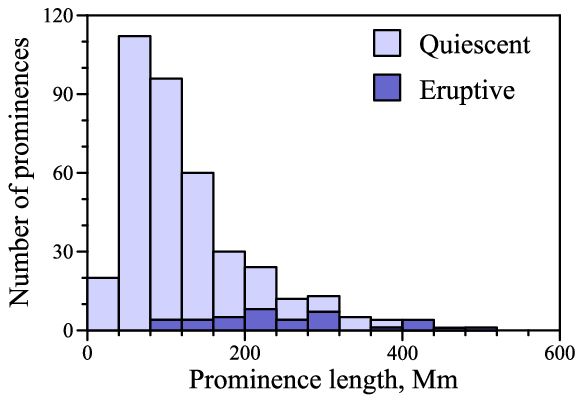}}
	\subfloat[\vspace{-2mm}]{\label{fig:distrib_hgt}\includegraphics[]{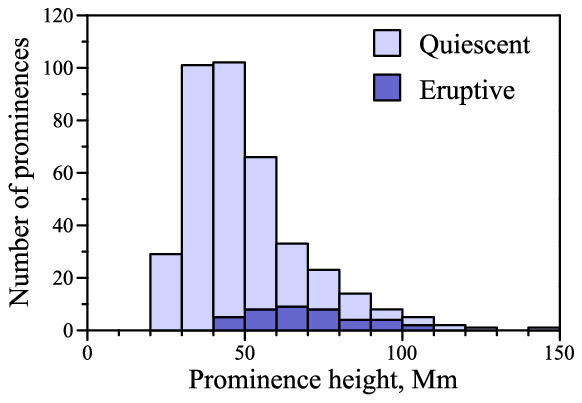}}
	\\ \vspace{-4mm}
	\subfloat[\vspace{-2mm}]{\label{fig:distrib_ltoh}\includegraphics[]{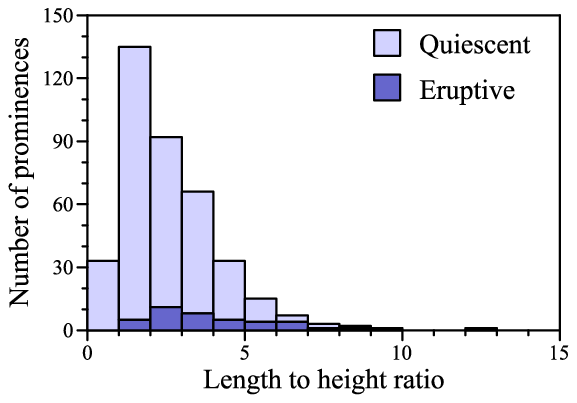}}
	\subfloat[\vspace{-2mm}]{\label{fig:distrib_lat}\includegraphics[]{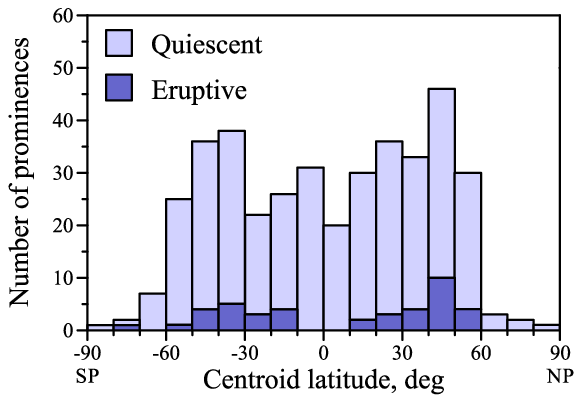}}
	\\ \vspace{-4mm}
	\subfloat[\vspace{-2mm}]{\label{fig:distrib_vel}\includegraphics[]{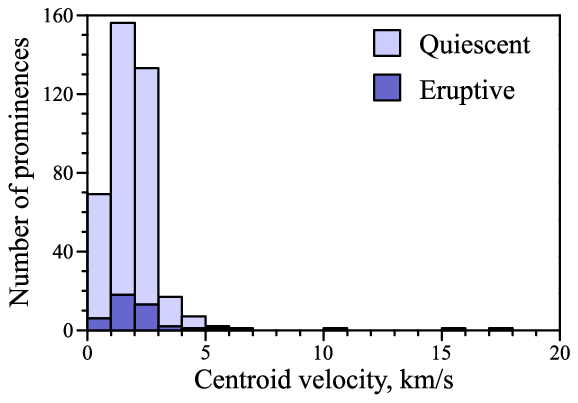}}
	\subfloat[\vspace{-2mm}]{\label{fig:distrib_ltime}\includegraphics[]{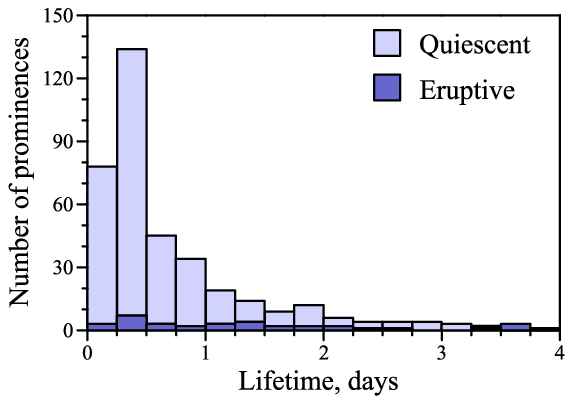}}
	\caption{\protect\subref{fig:distrib_len} Distribution of the lengths of prominences. \protect\subref{fig:distrib_hgt} Distribution of the heights above solar limb. \protect\subref{fig:distrib_ltoh} Distribution of the shapes in terms of length to height ratio. \protect\subref{fig:distrib_lat} Latitudinal distribution of prominences; negative latitudes correspond to the southern hemisphere, positive to the northern hemisphere \protect\subref{fig:distrib_vel} Distribution of centroid velocities. \protect\subref{fig:distrib_ltime} Distribution of observed lifetimes. The parameters in panels a--e are averaged over the prominence lifetime.}
\end{figure}

For the period of four months, we have detected a total of 389 quiescent prominences, 41 of them were classified as eruptive.
Detected prominences vary largely in size, the biggest is 500~Mm long and 150~Mm high and the smallest 25~Mm both in length and height (Figures~\ref{fig:distrib_len},\subref*{fig:distrib_hgt}).
Thus, the characteristic size of a prominence $\widetilde{L} = \sqrt{LH}$ (where $H$ is the prominence height and $L$ is the prominence length along the limb) is less than 70~Mm for 50~\% of the prominences.
Most of the detected prominences are stretched along the limb; for only 10~\% the ratio $L/H$ was found to be less than 1 (Figure~\ref{fig:distrib_ltoh}).
As for the latitudinal distribution of prominences, more than 95~\% are concentrated at latitudes below $\pm 60^{\circ}$, the maximal probability density falls at medium latitudes around $\pm40^{\circ}$ (Figure~\ref{fig:distrib_lat}).

Most of the detected prominences proved to be stable, showing practically no apparent bulk motion.
This is well demonstrated by Figure~\ref{fig:distrib_vel}; for more than 92~\% of the prominences the centroid velocity, averaged over the prominence lifetime, does not exceed 3~km~s$^{-1}$.
This value, however, does not reveal the inner plasma motions in prominences.
One of the surprising results is that around 75~\% of the prominences were detected off-limb for less than 24 hours (Figure~\ref{fig:distrib_ltime}).
Manual investigation has shown that smaller prominences tend to be more dynamic than the larger ones, and often disappear from the limb much faster than what the solar rotation would imply.
In addition, it was found that eruptive prominences show in general the same distribution of the inferred parameters, except for the fact that they tend to be among the bigger prominences.
This suggests the existence of a certain critical mass, which should be reached before a prominence can become unstable.

\subsection{Estimation Models}\label{sect:models}

In contrast to determining spatial and dynamic characteristics, estimation of the mass and, consequently, the gravitational energy of prominences proves to be a much more difficult task.
The most precise methods developed so far are generally based on measuring the fraction of background radiation absorbed by a prominence, most commonly in the iron EUV lines \citep{2005ApJ...618..524G, 2012ASPC..454..117S}.
Unfortunately, this approach can be only used in a limited number of cases, when prominences are observed against a sufficiently uniform background, which does not change much over time.
We have attempted, however, to apply this technique using the 171~\AA\ channel data of TESIS, but as  expected, with no promising result. 
Interesting developments to this method have been proposed by \cite{2013ApJ...772...71L} and \cite{2013ApJ...764..165W}.
These authors employed observations in several EUV lines with different absorption coefficients in order to exclude redundant parameters and thus to achieve higher precisions.
Both these techniques, however, require prominence material to be cool enough so that it would not emit in the \ion{He}{ii} 304~\AA\ line.
\cite{2003ApJ...594.1060L} have also proposed to estimate a prominence mass as the minimal amount of plasma needed to maintain stable the current magnetic configuration; evidently, this approach is not applicable to our case either.
Thus, we have no better option than to limit our study with only rough estimates of prominence masses and energies.
Though being inapplicable for the detailed studies of prominences, such estimates, however, can be fairly useful for investigating the general properties of a substantial number of events.

In our attempt to estimate the mass of a prominence, we rely on the fact that the prominence plasma proves to be optically thick when observed in the \ion{He}{ii} 304~\AA\ line.
This implies that the observed radiation is mainly formed in an outer layer of thickness $L_0$, where the radiation from the underlying layers of the prominence is absorbed by helium ions. From the Beer-Lambert law, the thickness $L_0$ is related to the mean number density of partly ionized helium atoms $n_{Heii}$ as
\begin{equation}
L_0 \sim 1/\varepsilon \sim 1/n_{Heii} \,,
\end{equation}
where $\varepsilon$ is the mean absorption coefficient of the emitting layer along the line of sight.
We therefore assume that this outer layer is isothermal (with temperature $T_0$) and homogeneous along the line of sight, which means that the observed flux $F$ is given by
\begin{equation}
F = G(T_0)EM(T_0) \sim n_{Heii}n_e L_0 \sim n_e \,,
\label{eq:rad_model}
\end{equation}
where $n_e$ is the electron density,  $G$ is the contribution function and $EM$ is the emission measure.
Numerical simulations by \cite{2009A&A...503..663G}, which consider a much more realistic radiation transfer, including scattering of incident radiation from the Sun, imply a more complex dependence.
However, for typical plasma temperatures and densities in prominences these simulations suggest that a linear dependence as in Equation~(\ref{eq:rad_model}) is an acceptable approximation.

In the simplest case, one can relate the volume of a prominence, $V$, to its observed area, $A$, as $V \sim A^{3/2}$.
Here, we use an alternative method instead, and estimate the thickness of a prominence, $D$, at each pixel of the detected area assuming that prominences have a more or less round section.
This means that the prominence attains its maximum thickness $D_{max}$ at points that are the most distant from the boundaries and that $D_{max}$ is equal to twice that distance.
For the rest of the pixels, we calculate the thickness $D$ as:
\begin{equation}
%D=2\left[D_{max}^2 - \left( D_{max}-R \right)^2\right]^{1/2} \,,
D = 2\, \sqrt{ D_{max}^2 - \left( D_{max}-R \right)^2 } \,,
\end{equation}
where $R$ is the minimal distance from the given point to the prominence edge.
However, during our study, we have found that for the set of detected prominences this method gives a clear dependence $V \sim A^{1.33\pm0.07}$, which is more consistent with the fact that prominences tend to be elongated rather than spherical in shape.
Both methods, however, are far from being precise as they do not take into account projection effects and, again, can be used for a rough evaluation only.

\begin{figure}[t]
	\centering
	\subfloat[\vspace{-2mm}]{\label{fig:tot_mass}\includegraphics[]{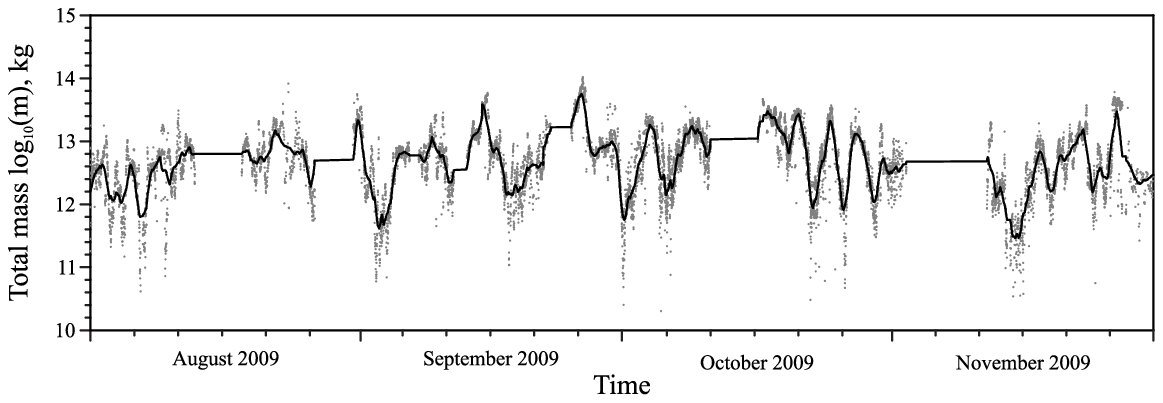}}
	\\ \vspace{-4mm}
	\subfloat[\vspace{-2mm}]{\label{fig:tot_en}\includegraphics[]{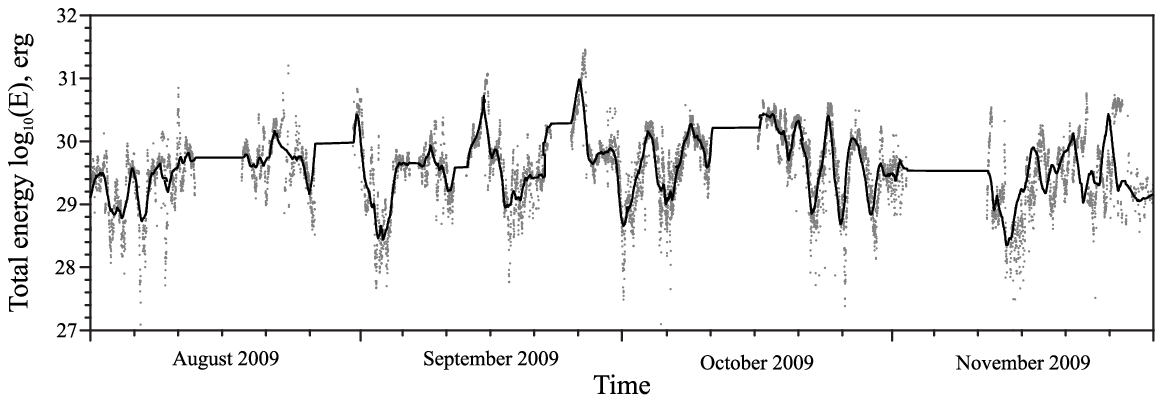}}
	\caption{Total mass \protect\subref{fig:tot_mass} and gravitational energy \protect\subref{fig:tot_en} of prominences simultaneously observed in a single image as they vary through the observation period. Grey dots correspond to individual images and the  black line is the one-day smoothed trend.}
\end{figure}

These two models allow us to obtain estimates of the mass and, therefore, of the gravitational energy of a detected prominence.
We choose the proportionality coefficient in Equation~(\ref{eq:rad_model}) such that, for all the detected prominences, the mean value of the electron density would equal a typical value of $2\times10^{10}$~cm$^{-3}$, which was found in previous studies of individual prominences \citep{1986A&A...156...90B, 1994SoPh..154..231B, 1994scs..conf..381W}.
We have to assume here, of course, that the electron density inside a prominence remains proportional to that in its outer shell along the line of sight.
Earlier studies of  prominences have shown that their ionisation degree is typically in the range of 0.5--0.8 \citep{1980ApJ...240..908O,1981SoPh...69..313K} and their helium abundance is around 0.1 \citep{1978ApJ...221..677H}.
Therefore, suggesting that most of the prominence mass is due to the hydrogen atoms, and the low helium abundance will not imply a large error, we calculate the mean density of the prominence plasma by multiplying the inferred electron density by the proton mass.
Subsequently, multiplying this plasma density by the previously calculated thickness of the prominence, we obtain the column mass at the every point of its visible surface.
Thus, we can obtain the total mass of a prominence by integrating these column masses over the observed area of a prominence, and the gravitational energy by integrating the column masses multiplied by the gravitational potential (here, we assume that the detected prominences are in the plane of the sky).
This is, by no means, a precise method, but with the data available, it still can be used for rough estimates of these important values.

\subsection{Mass and Gravitational Energy}

Having summed up masses and energies of all individual prominences detected in a single image, we can calculate the total mass of prominences observed simultaneously off-limb.
During the observation period, the total mass of the prominences varied greatly with time between $10^{12,2}$ and $10^{13.6}$~kg (Figure~\ref{fig:tot_mass}), as well as the total gravitational energy, which was found to be in the range from $10^{28.7}$ up to $10^{30.4}$~erg (Figure~\ref{fig:tot_en}).

\begin{figure}[t]
	\centering
	\includegraphics[]{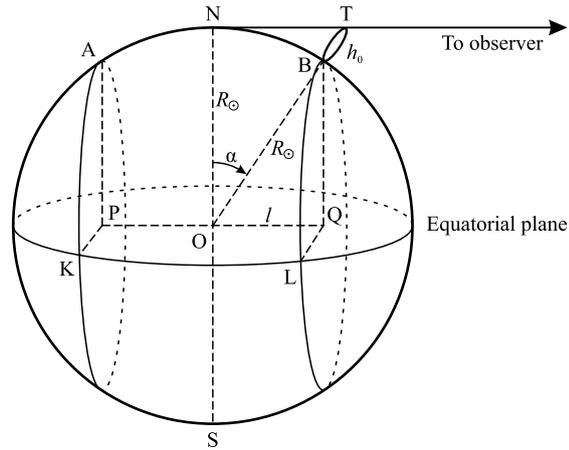} \vspace{1mm}
	\caption{Geometrical model used to evaluate the fraction of prominences visible off-limb. The prominence BT is located at a critical position, where it may become not visible above the limb.}
	\label{fig:geometry}
\end{figure}

In effect, we can observe, and therefore detect, only a certain fraction of all prominences present simultaneously on the Sun, which are located close to the limb and are high enough to be seen off-limb at their position.
To estimate this fraction, we develop a simple geometrical model presented in Figure~\ref{fig:geometry}.
Here, a prominence BT, which has a typical height $h_0$, is located at a critical position where it becomes no more visible above the limb.
We can assume $h_0\simeq40$~Mm, taking the most probable value from the distribution in Figure~\ref{fig:distrib_hgt}; however, since prominences are not visible below $\sim15$~Mm because of strong background radiation from the EUV spicules, we take here $h_0=25$~Mm and the solar radius $R_{\odot}=715$~Mm.
According to this model, we are able to observe only those prominences that are located between the planes APK and BQL (Figure~\ref{fig:geometry}).
The area of the solar surface confined between these two planes is given by:
\begin{equation}
A_{vis} = 2\pi R_{\odot} 2l = 4\pi R_{\odot}^2 \sin\alpha \,,
\end{equation}
where $2l=$~PQ and $\alpha$ is the $\widehat{\mathrm{NOB}}$. Since $4\pi R_{\odot}^2$ is the total surface area of the Sun, we can observe off-limb a fraction of around  
\begin{equation}
\sin\alpha = \sqrt{ 1 - R_{\odot}^2 / \left( R_{\odot} + h_0 \right)^2 } \simeq \sqrt{\vphantom\sum  2h_0 / R_{\odot}} \simeq 0.25
\end{equation}
of all prominences present anywhere on the Sun, including those visible as filaments on the disk.
Taking this into account, we conclude that at solar minimum, all quiet-Sun prominences contain simultaneously about $10^{12}$--$10^{14}$~kg of plasma and their total gravitational energy is of the order of $10^{29}$--$10^{31}$~erg.

\begin{figure}[t]
	\centering \vspace{-4mm}
	\subfloat[\vspace{-2mm}]{\label{fig:distrib_en}\includegraphics[]{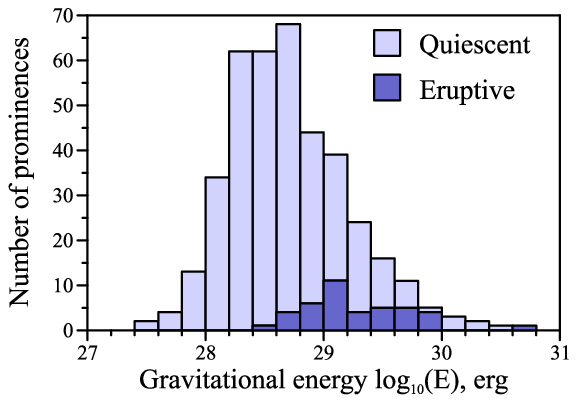}}
	\subfloat[\vspace{-2mm}]{\label{fig:loglog_en}\includegraphics[]{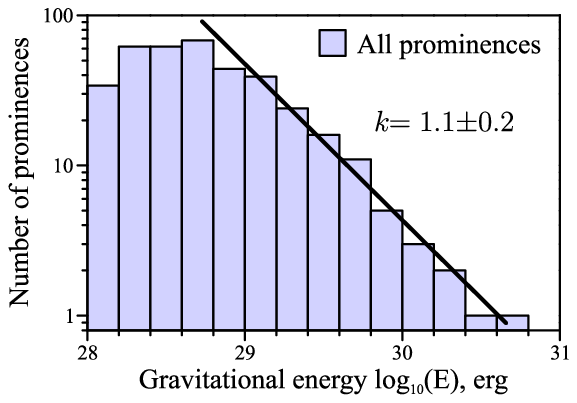}}
	\caption{\protect\subref{fig:distrib_en} Gravitational energy spectrum of the prominences. \protect\subref{fig:loglog_en} The slope of the same distribution for a limited energy range in a log-log scale.}
\end{figure}

The big number of detected events enables us to investigate the form of their gravitational energy spectrum, which is shown in Figure~\ref{fig:distrib_en}.
Although this spectrum may be shifted towards lower or higher energies due to the arbitrary choice of the coefficient in Equation~(\ref{eq:rad_model}), its shape is only biased by our assumptions about the prominence morphology and density distribution along the line of sight.
The main feature of this spectrum is that it has a clear maximum at around $10^{28.6}$~erg.
We believe that this behaviour can be explained by the existence of a low sensitivity threshold in our algorithm.
Most likely, we are not able to detect small and low-lying prominences in the 304~\AA\ images, as they become barely distinguishable from the strong background below 15~Mm (Figure~\ref{fig:rad_prof}).
Indeed, we have detected no prominences lower than 20~Mm (Figure~\ref{fig:distrib_hgt}).
On the other hand, there can be no such small prominences at all, as they may become unstable at these spatial scales or indistinguishable from smaller objects seen above the limb.
If so, the conventional definition of a prominence becomes no more applicable in this case.
Anyway, we consider only that part of the spectrum above $10^{28.6}$~erg to have a physical meaning.
In a log-log plot (Figure~\ref{fig:loglog_en}), it shows a clear power-law dependence, its index given by the slope of the distribution that we find to be $-1.1 \pm 0.2$.
Hence, in this case, the power-law index is close to $-1$, which means, that at least within the sensitivity range of our method, the gravitational energy is close to be evenly distributed among prominences of different sizes.

\section{Discussion}\label{sect:discussion}

As we mentioned above, most of the recent prominence studies are limited to particular cases, in general to the biggest events, which are easier to identify by eye.
At the same time, the whole set of prominences is of great interest, and several questions remain, to which answers cannot be given by studies of individual events.
It is unclear, for example, how the physical properties of prominences vary from one to the other and what their long-term behaviour is, \textit{i.e.}, how they change throughout the solar cycle (for a discussion, see a review by \citealp{2002SoPh..208..253P}).
Detailed catalogues that give not only qualitative, but also quantitative information and comprise a significant number of their measurable parameters would enable us to perform large statistical studies of these phenomena.
These studies, perhaps being not as precise as the individual ones, allow us, however, to cover a much wider range of events, maintaining a consistent point of view, less dependent on the selection of the objects examined.

A catalogue of prominences above the limb would also be an appropriate complement to the existing catalogues of filaments on disk \citep{2005SoPh..228..361Z}, as they show prominences in two orthogonal projections.
Equally promising are combined analysis of prominence catalogues with those of associated events in the solar corona.
For example, deeper conclusions can be achieved  by combining prominence catalogues and existing CME catalogues \citep{2009EM&P..104..295G, 2009ApJ...691.1222R, 2013SoPh..288..269F}.
They are both observed in the direction orthogonal to the line of sight, and comparative studies of a great number of events would help to advance our understanding of the mechanisms underlying CME initiation.
A joint study including flare catalogues \citep{2012SoPh..279..317W, 2014SoPh..289..919A} would be of no less interest, as prominence eruptions are often an essential part these complex processes.

Within the limits of this study, we have investigated a relatively short period.
Nevertheless, it provides a lot of new information about general properties of the entire set of prominences, including rather small ones.
One should also note that this observation period coincides with the end of an extraordinarily long solar minimum, which is a unique event by itself.
The main advantage of the algorithm developed is a well-founded technique of background removal and a more accurate reconstruction of prominences, which increases the reliability of the measured prominence properties.
Moreover, this algorithm is able to discriminate between the prominence and off-limb manifestations of active regions without the need of learning the method previously or using multi-wavelength observations.

The obtained values of mass and gravitational energy of prominences, despite the simplicity of the evaluation method used, are, nevertheless, the first attempt to determine these important properties for a substantial number of prominences.
These data are useful to constrain existing models of prominence formation and evolution and to study their role in the general mechanisms of mass and energy transfer in the solar corona.
It is also relevant that these properties are not only obtained for the largest events, but for the whole ensemble.
Therefore, they give an early insight of how these parameters are distributed among prominences of different size.

Because of the relatively short operation time of TESIS, our next goal is to adapt our algorithm to process the data obtained by other instruments, primarily those on board SDO.
These data will permit us to examine prominences on much larger time scales and thus to detect long-term variations of their properties throughout the solar cycle, if any, and to find out why this solar minimum was so different from the others.
For now, we can see two major problems that we will encounter in this endeavour.
First, our study did not include active prominences, as they were not numerous during the examined period.
Closer to the solar maximum, however, the number of this type of prominences will greatly increase along with the general increase of solar activity.
Nevertheless, we believe that our algorithm can be successfully used to detect these prominences as well, against the enhanced active region background.
The other problem is that our method of determining the background becomes inapplicable because of a more complex morphology of the corona during the rest of the solar cycle.
Anyhow, the statistical approach used here remains a powerful tool to deal with this issue as well.

\begin{acks}
This work was supported by the Russian Foundation for Basic Research (grant 14-02-00945) and by the Program \textnumero~9 for fundamental research of the Praesidium of the Russian Academy of Sciences.
We are grateful to all members of the TESIS development team who have put much effort in the design and operation of this instrument.

\end{acks}

\bibliographystyle{spr-mp-sola}
\bibliography{promstat}

\begin{thebibliography}{44}
% BibTex style file: spr-mp-sola.bst (nameyear), 2014-02-13
\ifx\bisbn     \undefined \def\bisbn  #1{ISBN #1}\fi
\ifx\binits    \undefined \def\binits#1{#1}\fi
\ifx\bauthor   \undefined \def\bauthor#1{#1}\fi
\ifx\batitle   \undefined \def\batitle#1{#1}\fi
\ifx\bjtitle   \undefined \def\bjtitle#1{\textit{#1}}\fi
\ifx\bvolume   \undefined \def\bvolume#1{\textbf{#1}}\fi
\ifx\byear     \undefined \def\byear#1{#1}\fi
\ifx\bissue    \undefined \def\bissue#1{#1}\fi
\ifx\bfpage    \undefined \def\bfpage#1{#1}\fi
\ifx\blpage    \undefined \def\blpage #1{#1}\fi
\ifx\burl      \undefined \def\burl#1{\textsf{#1}}\fi
\ifx\href      \undefined \def\href#1#2{\textsf{#2}}\fi
\ifx\betal     \undefined \def\betal{\textit{et al.}}\fi
\ifx\bctitle   \undefined \def\bctitle#1{#1}\fi
\ifx\beditor   \undefined \def\beditor#1{#1}\fi
\ifx\bbtitle   \undefined \def\bbtitle#1{\textit{#1}}\fi
\ifx\bedition  \undefined \def\bedition#1{#1}\fi
\ifx\bseriesno \undefined \def\bseriesno#1{\textbf{#1}}\fi
\ifx\blocation \undefined \def\blocation#1{#1}\fi
\ifx\bsertitle \undefined \def\bsertitle#1{\textit{#1}}\fi
\ifx\bsnm      \undefined \def\bsnm#1{#1}\fi
\ifx\bsuffix   \undefined \def\bsuffix#1{#1}\fi
\ifx\bparticle \undefined \def\bparticle#1{#1}\fi
\ifx\barticle  \undefined \def\barticle#1{}\fi
\ifx\binstitute  \undefined \def\binstitute#1{#1}\fi
\ifx\bpublisher  \undefined \def\bpublisher#1{#1}\fi
\ifx\doiurl    \undefined
  \def\doiurl#1{\href{http://dx.doi.org/#1}{\textsf{DOI}}}\fi
\ifx\arxivurl  \undefined
  \def\arxivurl#1{\href{http://arxiv.org/abs/#1}{\textsf{arXiv}}}\fi
\ifx\adsurl    \undefined
  \def\adsurl#1{\href{http://adsabs.harvard.edu/abs/#1}{\textsf{ADS}}}\fi
\ifx\botherref \undefined \def\botherref#1{}\fi
\ifx\url       \undefined \def\url#1{\textsf{#1}}\fi
\ifx\bchapter  \undefined \def\bchapter#1{}\fi
\ifx\bbook     \undefined \def\bbook#1{}\fi
\ifx\bcomment  \undefined \def\bcomment#1{#1}\fi
\ifx\oauthor   \undefined \def\oauthor#1{#1}\fi
\ifx\citeauthoryear \undefined\def \citeauthoryear#1{#1}\fi
\def\endbibitem {}
\ifx\bconflocation  \undefined \def\bconflocation#1{#1} \fi

\bibitem[\protect\citeauthoryear{{Aschwanden}
  \textit{et~al.}}{2014}]{2014SoPh..289..919A}
\begin{barticle}
\bauthor{\bsnm{{Aschwanden}}, \binits{M.J.}},
\bauthor{\bsnm{{W{\"u}lser}}, \binits{J.-P.}},
\bauthor{\bsnm{{Nitta}}, \binits{N.V.}},
\bauthor{\bsnm{{Lemen}}, \binits{J.R.}},
\bauthor{\bsnm{{Freeland}}, \binits{S.}},
\bauthor{\bsnm{{Thompson}}, \binits{W.T.}}:
\byear{2014},
\batitle{{STEREO/ Extreme Ultraviolet Imager (EUVI) Event Catalog 2006--2012}}.
\bjtitle{\solphys}
\bvolume{289},
\bfpage{919}.
\doiurl{10.1007/s11207-013-0378-5}.
\adsurl{2014SoPh..289..919A}.
\end{barticle}
\endbibitem

\bibitem[\protect\citeauthoryear{{Bommier}, {Leroy}, and
  {Sahal-Brechot}}{1986}]{1986A&A...156...90B}
\begin{barticle}
\bauthor{\bsnm{{Bommier}}, \binits{V.}},
\bauthor{\bsnm{{Leroy}}, \binits{J.L.}},
\bauthor{\bsnm{{Sahal-Brechot}}, \binits{S.}}:
\byear{1986},
\batitle{{The Linear Polarization of Hydrogen H-Beta Radiation and the Joint
  Diagnostic of Magnetic Field Vector and Electron Density in Quiescent
  Prominences --- Part Two --- the Electron Density}}.
\bjtitle{\aap}
\bvolume{156},
\bfpage{90}.
\adsurl{1986A\%26A...156...90B}.
\end{barticle}
\endbibitem

\bibitem[\protect\citeauthoryear{{Bommier}
  \textit{et~al.}}{1994}]{1994SoPh..154..231B}
\begin{barticle}
\bauthor{\bsnm{{Bommier}}, \binits{V.}},
\bauthor{\bsnm{{Landi Degl'Innocenti}}, \binits{E.}},
\bauthor{\bsnm{{Leroy}}, \binits{J.-L.}},
\bauthor{\bsnm{{Sahal-Brechot}}, \binits{S.}}:
\byear{1994},
\batitle{{Complete Determination of the Magnetic Field Vector and of the
  Electron Density in 14 Prominences from Linear Polarization Measurements in
  the HeI D3 and H-Alpha Lines}}.
\bjtitle{\solphys}
\bvolume{154},
\bfpage{231}.
\doiurl{10.1007/BF00681098}.
\adsurl{1994SoPh..154..231B}.
\end{barticle}
\endbibitem

\bibitem[\protect\citeauthoryear{{Cushman} and
  {Rense}}{1978}]{1978SoPh...58..299C}
\begin{barticle}
\bauthor{\bsnm{{Cushman}}, \binits{G.W.}},
\bauthor{\bsnm{{Rense}}, \binits{W.A.}}:
\byear{1978},
\batitle{{Solar \ion{He}{ii} 304~\AA\ and \ion{SI}{xi} 303~\AA\ line
  profiles}}.
\bjtitle{\solphys}
\bvolume{58},
\bfpage{299}.
\doiurl{10.1007/BF00157275}.
\adsurl{1978SoPh...58..299C}.
\end{barticle}
\endbibitem

\bibitem[\protect\citeauthoryear{{Filippov} and
  {Koutchmy}}{2008}]{2008AnGeo..26.3025F}
\begin{barticle}
\bauthor{\bsnm{{Filippov}}, \binits{B.}},
\bauthor{\bsnm{{Koutchmy}}, \binits{S.}}:
\byear{2008},
\batitle{{Causal relationships between eruptive prominences and coronal mass
  ejections}}.
\bjtitle{Ann. Geophys.}
\bvolume{26},
\bfpage{3025}.
\doiurl{10.5194/angeo-26-3025-2008}.
\adsurl{2008AnGeo..26.3025F}.
\end{barticle}
\endbibitem

\bibitem[\protect\citeauthoryear{{Floyd}
  \textit{et~al.}}{2013}]{2013SoPh..288..269F}
\begin{barticle}
\bauthor{\bsnm{{Floyd}}, \binits{O.}},
\bauthor{\bsnm{{Lamy}}, \binits{P.}},
\bauthor{\bsnm{{Boursier}}, \binits{Y.}},
\bauthor{\bsnm{{Llebaria}}, \binits{A.}}:
\byear{2013},
\batitle{{ARTEMIS II: A Second-Generation Catalog of LASCO Coronal Mass
  Ejections Including Mass and Kinetic Energy}}.
\bjtitle{\solphys}
\bvolume{288},
\bfpage{269}.
\doiurl{10.1007/s11207-013-0281-0}.
\adsurl{2013SoPh..288..269F}.
\end{barticle}
\endbibitem

\bibitem[\protect\citeauthoryear{{Foullon} and
  {Verwichte}}{2006}]{2006SoPh..234..135F}
\begin{barticle}
\bauthor{\bsnm{{Foullon}}, \binits{C.}},
\bauthor{\bsnm{{Verwichte}}, \binits{E.}}:
\byear{2006},
\batitle{{Automated Detection of EUV Prominences}}.
\bjtitle{\solphys}
\bvolume{234},
\bfpage{135}.
\doiurl{10.1007/s11207-006-0054-0}.
\adsurl{2006SoPh..234..135F}.
\end{barticle}
\endbibitem

\bibitem[\protect\citeauthoryear{{Galsgaard} and
  {Longbottom}}{1999}]{1999ApJ...510..444G}
\begin{barticle}
\bauthor{\bsnm{{Galsgaard}}, \binits{K.}},
\bauthor{\bsnm{{Longbottom}}, \binits{A.W.}}:
\byear{1999},
\batitle{{Formation of Solar Prominences by Flux Convergence}}.
\bjtitle{\apj}
\bvolume{510},
\bfpage{444}.
\doiurl{10.1086/306559}.
\adsurl{1999ApJ...510..444G}.
\end{barticle}
\endbibitem

\bibitem[\protect\citeauthoryear{{Gilbert}, {Holzer}, and
  {MacQueen}}{2005}]{2005ApJ...618..524G}
\begin{barticle}
\bauthor{\bsnm{{Gilbert}}, \binits{H.R.}},
\bauthor{\bsnm{{Holzer}}, \binits{T.E.}},
\bauthor{\bsnm{{MacQueen}}, \binits{R.M.}}:
\byear{2005},
\batitle{{A New Technique for Deriving Prominence Mass from SOHO/EIT Fe XII
  (19.5 Nanometers) Absorption Features}}.
\bjtitle{\apj}
\bvolume{618},
\bfpage{524}.
\doiurl{10.1086/425975}.
\adsurl{2005ApJ...618..524G}.
\end{barticle}
\endbibitem

\bibitem[\protect\citeauthoryear{{Gilbert}
  \textit{et~al.}}{2000}]{2000ApJ...537..503G}
\begin{barticle}
\bauthor{\bsnm{{Gilbert}}, \binits{H.R.}},
\bauthor{\bsnm{{Holzer}}, \binits{T.E.}},
\bauthor{\bsnm{{Burkepile}}, \binits{J.T.}},
\bauthor{\bsnm{{Hundhausen}}, \binits{A.J.}}:
\byear{2000},
\batitle{{Active and Eruptive Prominences and Their Relationship to Coronal
  Mass Ejections}}.
\bjtitle{\apj}
\bvolume{537},
\bfpage{503}.
\doiurl{10.1086/309030}.
\adsurl{2000ApJ...537..503G}.
\end{barticle}
\endbibitem

\bibitem[\protect\citeauthoryear{{Gopalswamy}
  \textit{et~al.}}{2003}]{2003ApJ...586..562G}
\begin{barticle}
\bauthor{\bsnm{{Gopalswamy}}, \binits{N.}},
\bauthor{\bsnm{{Shimojo}}, \binits{M.}},
\bauthor{\bsnm{{Lu}}, \binits{W.}},
\bauthor{\bsnm{{Yashiro}}, \binits{S.}},
\bauthor{\bsnm{{Shibasaki}}, \binits{K.}},
\bauthor{\bsnm{{Howard}}, \binits{R.A.}}:
\byear{2003},
\batitle{{Prominence Eruptions and Coronal Mass Ejection: A Statistical Study
  Using Microwave Observations}}.
\bjtitle{\apj}
\bvolume{586},
\bfpage{562}.
\doiurl{10.1086/367614}.
\adsurl{2003ApJ...586..562G}.
\end{barticle}
\endbibitem

\bibitem[\protect\citeauthoryear{{Gopalswamy}
  \textit{et~al.}}{2009}]{2009EM&P..104..295G}
\begin{barticle}
\bauthor{\bsnm{{Gopalswamy}}, \binits{N.}},
\bauthor{\bsnm{{Yashiro}}, \binits{S.}},
\bauthor{\bsnm{{Michalek}}, \binits{G.}},
\bauthor{\bsnm{{Stenborg}}, \binits{G.}},
\bauthor{\bsnm{{Vourlidas}}, \binits{A.}},
\bauthor{\bsnm{{Freeland}}, \binits{S.}},
\bauthor{\bsnm{{Howard}}, \binits{R.}}:
\byear{2009},
\batitle{{The SOHO/LASCO CME Catalog}}.
\bjtitle{Earth Moon Planets}
\bvolume{104},
\bfpage{295}.
\doiurl{10.1007/s11038-008-9282-7}.
\adsurl{2009EM\%26P..104..295G}.
\end{barticle}
\endbibitem

\bibitem[\protect\citeauthoryear{{Gouttebroze} and
  {Labrosse}}{2009}]{2009A&A...503..663G}
\begin{barticle}
\bauthor{\bsnm{{Gouttebroze}}, \binits{P.}},
\bauthor{\bsnm{{Labrosse}}, \binits{N.}}:
\byear{2009},
\batitle{{Radiative transfer in cylindrical threads with incident radiation.
  VI. A hydrogen plus helium system}}.
\bjtitle{\aap}
\bvolume{503},
\bfpage{663}.
\doiurl{10.1051/0004-6361/200811483}.
\adsurl{2009A\%26A...503..663G}.
\end{barticle}
\endbibitem

\bibitem[\protect\citeauthoryear{{Heasley} and
  {Milkey}}{1978}]{1978ApJ...221..677H}
\begin{barticle}
\bauthor{\bsnm{{Heasley}}, \binits{J.N.}},
\bauthor{\bsnm{{Milkey}}, \binits{R.W.}}:
\byear{1978},
\batitle{{Structure and spectrum of quiescent prominences. III - Application of
  theoretical models in helium abundance determinations}}.
\bjtitle{\apj}
\bvolume{221},
\bfpage{677}.
\doiurl{10.1086/156072}.
\adsurl{1978ApJ...221..677H}.
\end{barticle}
\endbibitem

\bibitem[\protect\citeauthoryear{{Heinzel}}{2007}]{2007ASPC..370...46H}
\begin{bchapter}
\bauthor{\bsnm{{Heinzel}}, \binits{P.}}:
\byear{2007},
\bctitle{{Multiwavelength Observations of Solar Prominences}}.
In: \beditor{\bsnm{{Demircan}}, \binits{O.}},
\beditor{\bsnm{{Selam}}, \binits{S.O.}},
\beditor{\bsnm{{Albayrak}}, \binits{B.}} (eds.)
\bbtitle{Solar and Stellar Physics Through Eclipses},
\bsertitle{Astron. Soc. Pacific C.S.}
\bseriesno{370},
\bfpage{46}.
\adsurl{2007ASPC..370...46H}.
\end{bchapter}
\endbibitem

\bibitem[\protect\citeauthoryear{{Hirayama}}{1985}]{1985SoPh..100..415H}
\begin{barticle}
\bauthor{\bsnm{{Hirayama}}, \binits{T.}}:
\byear{1985},
\batitle{{Modern Observations of Solar Prominences}}.
\bjtitle{\solphys}
\bvolume{100},
\bfpage{415}.
\doiurl{10.1007/BF00158439}.
\adsurl{1985SoPh..100..415H}.
\end{barticle}
\endbibitem

\bibitem[\protect\citeauthoryear{{Kanno}, {Withbroe}, and
  {Noyes}}{1981}]{1981SoPh...69..313K}
\begin{barticle}
\bauthor{\bsnm{{Kanno}}, \binits{M.}},
\bauthor{\bsnm{{Withbroe}}, \binits{G.L.}},
\bauthor{\bsnm{{Noyes}}, \binits{R.W.}}:
\byear{1981},
\batitle{{Analysis of extreme-ultraviolet spectroheliograms of solar
  prominences}}.
\bjtitle{\solphys}
\bvolume{69},
\bfpage{313}.
\doiurl{10.1007/BF00149997}.
\adsurl{1981SoPh...69..313K}.
\end{barticle}
\endbibitem

\bibitem[\protect\citeauthoryear{{Kuzin}
  \textit{et~al.}}{2009}]{2009AdSpR..43.1001K}
\begin{barticle}
\bauthor{\bsnm{{Kuzin}}, \binits{S.V.}},
\bauthor{\bsnm{{Bogachev}}, \binits{S.A.}},
\bauthor{\bsnm{{Zhitnik}}, \binits{I.A.}},
\bauthor{\bsnm{{Pertsov}}, \binits{A.A.}},
\bauthor{\bsnm{{Ignatiev}}, \binits{A.P.}},
\bauthor{\bsnm{{Mitrofanov}}, \binits{A.M.}},
\bauthor{\bsnm{{Slemzin}}, \binits{V.A.}},
\bauthor{\bsnm{{Shestov}}, \binits{S.V.}},
\bauthor{\bsnm{{Sukhodrev}}, \binits{N.K.}},
\bauthor{\bsnm{{Bugaenko}}, \binits{O.I.}}:
\byear{2009},
\batitle{{TESIS Experiment on EUV Imaging Spectroscopy of the Sun}}.
\bjtitle{Adv. Space Res.}
\bvolume{43},
\bfpage{1001}.
\doiurl{10.1016/j.asr.2008.10.021}.
\adsurl{2009AdSpR..43.1001K}.
\end{barticle}
\endbibitem

\bibitem[\protect\citeauthoryear{{Kuzin}
  \textit{et~al.}}{2011}]{2011SoSyR..45..162K}
\begin{barticle}
\bauthor{\bsnm{{Kuzin}}, \binits{S.V.}},
\bauthor{\bsnm{{Zhitnik}}, \binits{I.A.}},
\bauthor{\bsnm{{Shestov}}, \binits{S.V.}},
\bauthor{\bsnm{{Bogachev}}, \binits{S.A.}},
\bauthor{\bsnm{{Bugaenko}}, \binits{O.I.}},
\bauthor{\bsnm{{Ignat'ev}}, \binits{A.P.}},
\bauthor{\bsnm{{Pertsov}}, \binits{A.A.}},
\bauthor{\bsnm{{Ulyanov}}, \binits{A.S.}},
\bauthor{\bsnm{{Reva}}, \binits{A.A.}},
\bauthor{\bsnm{{Slemzin}}, \binits{V.A.}},
\bauthor{\bsnm{{Sukhodrev}}, \binits{N.K.}},
\bauthor{\bsnm{{Ivanov}}, \binits{Y.S.}},
\bauthor{\bsnm{{Goncharov}}, \binits{L.A.}},
\bauthor{\bsnm{{Mitrofanov}}, \binits{A.V.}},
\bauthor{\bsnm{{Popov}}, \binits{S.G.}},
\bauthor{\bsnm{{Shergina}}, \binits{T.A.}},
\bauthor{\bsnm{{Solov'ev}}, \binits{V.A.}},
\bauthor{\bsnm{{Oparin}}, \binits{S.N.}},
\bauthor{\bsnm{{Zykov}}, \binits{A.M.}}:
\byear{2011},
\batitle{{The TESIS experiment on the CORONAS-PHOTON spacecraft}}.
\bjtitle{Solar Sys. Res.}
\bvolume{45},
\bfpage{162}.
\doiurl{10.1134/S0038094611020110}.
\adsurl{2011SoSyR..45..162K}.
\end{barticle}
\endbibitem

\bibitem[\protect\citeauthoryear{{Labrosse} and
  {Gouttebroze}}{2001}]{2001A&A...380..323L}
\begin{barticle}
\bauthor{\bsnm{{Labrosse}}, \binits{N.}},
\bauthor{\bsnm{{Gouttebroze}}, \binits{P.}}:
\byear{2001},
\batitle{{Formation of Helium Spectrum in Solar Quiescent Prominences}}.
\bjtitle{\aap}
\bvolume{380},
\bfpage{323}.
\doiurl{10.1051/0004-6361:20011395}.
\adsurl{2001A\%26A...380..323L}.
\end{barticle}
\endbibitem

\bibitem[\protect\citeauthoryear{{Labrosse}, {Dalla}, and
  {Marshall}}{2010}]{2010SoPh..262..449L}
\begin{barticle}
\bauthor{\bsnm{{Labrosse}}, \binits{N.}},
\bauthor{\bsnm{{Dalla}}, \binits{S.}},
\bauthor{\bsnm{{Marshall}}, \binits{S.}}:
\byear{2010},
\batitle{{Automatic Detection of Limb Prominences in 304 {\AA} EUV Images}}.
\bjtitle{\solphys}
\bvolume{262},
\bfpage{449}.
\doiurl{10.1007/s11207-009-9492-9}.
\adsurl{2010SoPh..262..449L}.
\end{barticle}
\endbibitem

\bibitem[\protect\citeauthoryear{{Labrosse}
  \textit{et~al.}}{2010}]{2010SSRv..151..243L}
\begin{barticle}
\bauthor{\bsnm{{Labrosse}}, \binits{N.}},
\bauthor{\bsnm{{Heinzel}}, \binits{P.}},
\bauthor{\bsnm{{Vial}}, \binits{J.-C.}},
\bauthor{\bsnm{{Kucera}}, \binits{T.}},
\bauthor{\bsnm{{Parenti}}, \binits{S.}},
\bauthor{\bsnm{{Gun{\'a}r}}, \binits{S.}},
\bauthor{\bsnm{{Schmieder}}, \binits{B.}},
\bauthor{\bsnm{{Kilper}}, \binits{G.}}:
\byear{2010},
\batitle{{Physics of Solar Prominences: I---Spectral Diagnostics and Non-LTE
  Modelling}}.
\bjtitle{\ssr}
\bvolume{151},
\bfpage{243}.
\doiurl{10.1007/s11214-010-9630-6}.
\adsurl{2010SSRv..151..243L}.
\end{barticle}
\endbibitem

\bibitem[\protect\citeauthoryear{{Landi} and
  {Reale}}{2013}]{2013ApJ...772...71L}
\begin{barticle}
\bauthor{\bsnm{{Landi}}, \binits{E.}},
\bauthor{\bsnm{{Reale}}, \binits{F.}}:
\byear{2013},
\batitle{{Prominence Plasma Diagnostics through Extreme-ultraviolet
  Absorption}}.
\bjtitle{\apj}
\bvolume{772},
\bfpage{71}.
\doiurl{10.1088/0004-637X/772/1/71}.
\adsurl{2013ApJ...772...71L}.
\end{barticle}
\endbibitem

\bibitem[\protect\citeauthoryear{{Leroy}, {Bommier}, and
  {Sahal-Brechot}}{1984}]{1984A&A...131...33L}
\begin{barticle}
\bauthor{\bsnm{{Leroy}}, \binits{J.L.}},
\bauthor{\bsnm{{Bommier}}, \binits{V.}},
\bauthor{\bsnm{{Sahal-Brechot}}, \binits{S.}}:
\byear{1984},
\batitle{{New data on the magnetic structure of quiescent prominences}}.
\bjtitle{\aap}
\bvolume{131},
\bfpage{33}.
\adsurl{1984A\%26A...131...33L}.
\end{barticle}
\endbibitem

\bibitem[\protect\citeauthoryear{{Lin}, {Soon}, and
  {Baliunas}}{2003}]{2003NewAR..47...53L}
\begin{barticle}
\bauthor{\bsnm{{Lin}}, \binits{J.}},
\bauthor{\bsnm{{Soon}}, \binits{W.}},
\bauthor{\bsnm{{Baliunas}}, \binits{S.L.}}:
\byear{2003},
\batitle{{Theories of Solar Eruptions: a Review}}.
\bjtitle{\nar}
\bvolume{47},
\bfpage{53}.
\doiurl{10.1016/S1387-6473(02)00271-3}.
\adsurl{2003NewAR..47...53L}.
\end{barticle}
\endbibitem

\bibitem[\protect\citeauthoryear{{Low}, {Fong}, and
  {Fan}}{2003}]{2003ApJ...594.1060L}
\begin{barticle}
\bauthor{\bsnm{{Low}}, \binits{B.C.}},
\bauthor{\bsnm{{Fong}}, \binits{B.}},
\bauthor{\bsnm{{Fan}}, \binits{Y.}}:
\byear{2003},
\batitle{{The Mass of a Solar Quiescent Prominence}}.
\bjtitle{\apj}
\bvolume{594},
\bfpage{1060}.
\doiurl{10.1086/377042}.
\adsurl{2003ApJ...594.1060L}.
\end{barticle}
\endbibitem

\bibitem[\protect\citeauthoryear{{Mackay}
  \textit{et~al.}}{2010}]{2010SSRv..151..333M}
\begin{barticle}
\bauthor{\bsnm{{Mackay}}, \binits{D.H.}},
\bauthor{\bsnm{{Karpen}}, \binits{J.T.}},
\bauthor{\bsnm{{Ballester}}, \binits{J.L.}},
\bauthor{\bsnm{{Schmieder}}, \binits{B.}},
\bauthor{\bsnm{{Aulanier}}, \binits{G.}}:
\byear{2010},
\batitle{{Physics of Solar Prominences: II---Magnetic Structure and Dynamics}}.
\bjtitle{\ssr}
\bvolume{151},
\bfpage{333}.
\doiurl{10.1007/s11214-010-9628-0}.
\adsurl{2010SSRv..151..333M}.
\end{barticle}
\endbibitem

\bibitem[\protect\citeauthoryear{{Morgan}, {Habbal}, and
  {Woo}}{2006}]{2006SoPh..236..263M}
\begin{barticle}
\bauthor{\bsnm{{Morgan}}, \binits{H.}},
\bauthor{\bsnm{{Habbal}}, \binits{S.R.}},
\bauthor{\bsnm{{Woo}}, \binits{R.}}:
\byear{2006},
\batitle{{The Depiction of Coronal Structure in White-Light Images}}.
\bjtitle{\solphys}
\bvolume{236},
\bfpage{263}.
\doiurl{10.1007/s11207-006-0113-6}.
\adsurl{2006SoPh..236..263M}.
\end{barticle}
\endbibitem

\bibitem[\protect\citeauthoryear{{Orrall} and
  {Schmahl}}{1980}]{1980ApJ...240..908O}
\begin{barticle}
\bauthor{\bsnm{{Orrall}}, \binits{F.Q.}},
\bauthor{\bsnm{{Schmahl}}, \binits{E.J.}}:
\byear{1980},
\batitle{{The H I Lyman continuum in solar prominences and its interpretation
  in the presence of inhomogeneities}}.
\bjtitle{\apj}
\bvolume{240},
\bfpage{908}.
\doiurl{10.1086/158304}.
\adsurl{1980ApJ...240..908O}.
\end{barticle}
\endbibitem

\bibitem[\protect\citeauthoryear{{Patsourakos} and
  {Vial}}{2002}]{2002SoPh..208..253P}
\begin{barticle}
\bauthor{\bsnm{{Patsourakos}}, \binits{S.}},
\bauthor{\bsnm{{Vial}}, \binits{J.-C.}}:
\byear{2002},
\batitle{{Soho Contribution to Prominence Science}}.
\bjtitle{\solphys}
\bvolume{208},
\bfpage{253}.
\doiurl{10.1023/A:1020510120772}.
\adsurl{2002SoPh..208..253P}.
\end{barticle}
\endbibitem

\bibitem[\protect\citeauthoryear{{Pettit}}{1932}]{1932ApJ....76....9P}
\begin{barticle}
\bauthor{\bsnm{{Pettit}}, \binits{E.}}:
\byear{1932},
\batitle{{Characteristic Features of Solar Prominences}}.
\bjtitle{\apj}
\bvolume{76},
\bfpage{9}.
\doiurl{10.1086/143396}.
\adsurl{1932ApJ....76....9P}.
\end{barticle}
\endbibitem

\bibitem[\protect\citeauthoryear{{Robbrecht}, {Berghmans}, and {Van der
  Linden}}{2009}]{2009ApJ...691.1222R}
\begin{barticle}
\bauthor{\bsnm{{Robbrecht}}, \binits{E.}},
\bauthor{\bsnm{{Berghmans}}, \binits{D.}},
\bauthor{\bsnm{{Van der Linden}}, \binits{R.A.M.}}:
\byear{2009},
\batitle{{Automated LASCO CME Catalog for Solar Cycle 23: Are CMEs Scale
  Invariant?}}
\bjtitle{\apj}
\bvolume{691},
\bfpage{1222}.
\doiurl{10.1088/0004-637X/691/2/1222}.
\adsurl{2009ApJ...691.1222R}.
\end{barticle}
\endbibitem

\bibitem[\protect\citeauthoryear{{Schwartz}
  \textit{et~al.}}{2012}]{2012ASPC..454..117S}
\begin{bchapter}
\bauthor{\bsnm{{Schwartz}}, \binits{P.}},
\bauthor{\bsnm{{F{\'a}rn{\'{\i}}k}}, \binits{F.}},
\bauthor{\bsnm{{Heinzel}}, \binits{P.}},
\bauthor{\bsnm{{Kotr{\v c}}}, \binits{P.}},
\bauthor{\bsnm{{Anzer}}, \binits{U.}}:
\byear{2012},
\bctitle{{Mass of Solar Prominences Estimated from Multi-Wavelength Data}}.
In: \beditor{\bsnm{{Sekii}}, \binits{T.}},
\beditor{\bsnm{{Watanabe}}, \binits{T.}},
\beditor{\bsnm{{Sakurai}}, \binits{T.}} (eds.)
\bbtitle{Hinode-3: The 3rd Hinode Science Meeting},
\bsertitle{Astron. Soc. Pacific C.S.}
\bseriesno{454},
\bfpage{117}.
\adsurl{2012ASPC..454..117S}.
\end{bchapter}
\endbibitem

\bibitem[\protect\citeauthoryear{{Severny} and {Khokhlova}}{1953}]{1953severny}
\begin{barticle}
\bauthor{\bsnm{{Severny}}, \binits{A.B.}},
\bauthor{\bsnm{{Khokhlova}}, \binits{V.L.}}:
\byear{1953},
\batitle{{Study of Motion and Emission of Solar Prominences}}.
\bjtitle{Izv. Krymsk. Astrofiz. Obs.}
\bvolume{10},
\bfpage{9}.
\end{barticle}
\endbibitem

\bibitem[\protect\citeauthoryear{{Tandberg-Hanssen}}{1995}]{1995tandberg}
\begin{bbook}
\bauthor{\bsnm{{Tandberg-Hanssen}}, \binits{E.}}:
\byear{1995},
\bbtitle{{The Nature of Solar Prominences}},
\bsertitle{Astrophys. Space Sci. Lib.}
\bseriesno{199},
\bpublisher{Kluwer Academic Pub.},
\blocation{Dordrecht}.
\end{bbook}
\endbibitem

\bibitem[\protect\citeauthoryear{{Thompson} and
  {Brekke}}{2000}]{2000SoPh..195...45T}
\begin{barticle}
\bauthor{\bsnm{{Thompson}}, \binits{W.T.}},
\bauthor{\bsnm{{Brekke}}, \binits{P.}}:
\byear{2000},
\batitle{{EUV Full-Sun Imaged Spectral Atlas Using the SOHO Coronal Diagnostic
  Spectrometer}}.
\bjtitle{\solphys}
\bvolume{195},
\bfpage{45}.
\doiurl{10.1023/A:1005203001242}.
\adsurl{2000SoPh..195...45T}.
\end{barticle}
\endbibitem

\bibitem[\protect\citeauthoryear{{van~Ballegooijen} and
  {Martens}}{1989}]{1989ApJ...343..971V}
\begin{barticle}
\bauthor{\bsnm{{van~Ballegooijen}}, \binits{A.A.}},
\bauthor{\bsnm{{Martens}}, \binits{P.C.H.}}:
\byear{1989},
\batitle{{Formation and eruption of solar prominences}}.
\bjtitle{\apj}
\bvolume{343},
\bfpage{971}.
\doiurl{10.1086/167766}.
\adsurl{1989ApJ...343..971V}.
\end{barticle}
\endbibitem

\bibitem[\protect\citeauthoryear{{Wang}
  \textit{et~al.}}{2010}]{2010ApJ...717..973W}
\begin{barticle}
\bauthor{\bsnm{{Wang}}, \binits{Y.}},
\bauthor{\bsnm{{Cao}}, \binits{H.}},
\bauthor{\bsnm{{Chen}}, \binits{J.}},
\bauthor{\bsnm{{Zhang}}, \binits{T.}},
\bauthor{\bsnm{{Yu}}, \binits{S.}},
\bauthor{\bsnm{{Zheng}}, \binits{H.}},
\bauthor{\bsnm{{Shen}}, \binits{C.}},
\bauthor{\bsnm{{Zhang}}, \binits{J.}},
\bauthor{\bsnm{{Wang}}, \binits{S.}}:
\byear{2010},
\batitle{{Solar Limb Prominence Catcher and Tracker (SLIPCAT): An Automated
  System and its Preliminary Statistical Results}}.
\bjtitle{\apj}
\bvolume{717},
\bfpage{973}.
\doiurl{10.1088/0004-637X/717/2/973}.
\adsurl{2010ApJ...717..973W}.
\end{barticle}
\endbibitem

\bibitem[\protect\citeauthoryear{{Watanabe}, {Masuda}, and
  {Segawa}}{2012}]{2012SoPh..279..317W}
\begin{barticle}
\bauthor{\bsnm{{Watanabe}}, \binits{K.}},
\bauthor{\bsnm{{Masuda}}, \binits{S.}},
\bauthor{\bsnm{{Segawa}}, \binits{T.}}:
\byear{2012},
\batitle{{Hinode Flare Catalogue}}.
\bjtitle{\solphys}
\bvolume{279},
\bfpage{317}.
\doiurl{10.1007/s11207-012-9983-y}.
\adsurl{2012SoPh..279..317W}.
\end{barticle}
\endbibitem

\bibitem[\protect\citeauthoryear{{Wiik}, {Heinzel}, and
  {Schmieder}}{1994}]{1994scs..conf..381W}
\begin{bchapter}
\bauthor{\bsnm{{Wiik}}, \binits{J.E.}},
\bauthor{\bsnm{{Heinzel}}, \binits{P.}},
\bauthor{\bsnm{{Schmieder}}, \binits{B.}}:
\byear{1994},
\bctitle{{Electron Densities in Solar Prominences.}}
In: \beditor{\bsnm{{Rusin}}, \binits{V.}},
\beditor{\bsnm{{Heinzel}}, \binits{P.}},
\beditor{\bsnm{{Vial}}, \binits{J.-C.}} (eds.)
\bbtitle{IAU Coll. 144: Solar Coronal Structures},
\bfpage{381}.
\adsurl{1994scs..conf..381W}.
\end{bchapter}
\endbibitem

\bibitem[\protect\citeauthoryear{{Williams}, {Baker}, and {van
  Driel-Gesztelyi}}{2013}]{2013ApJ...764..165W}
\begin{barticle}
\bauthor{\bsnm{{Williams}}, \binits{D.R.}},
\bauthor{\bsnm{{Baker}}, \binits{D.}},
\bauthor{\bsnm{{van Driel-Gesztelyi}}, \binits{L.}}:
\byear{2013},
\batitle{{Mass Estimates of Rapidly Moving Prominence Material from
  High-cadence EUV Images}}.
\bjtitle{\apj}
\bvolume{764},
\bfpage{165}.
\doiurl{10.1088/0004-637X/764/2/165}.
\adsurl{2013ApJ...764..165W}.
\end{barticle}
\endbibitem

\bibitem[\protect\citeauthoryear{{Zhang}}{2013}]{2013ApJ...777...52Z}
\begin{barticle}
\bauthor{\bsnm{{Zhang}}, \binits{Y.Z.}}:
\byear{2013},
\batitle{{The Formation and Eruption of Solar Quiescent Prominences}}.
\bjtitle{\apj}
\bvolume{777},
\bfpage{52}.
\doiurl{10.1088/0004-637X/777/1/52}.
\adsurl{2013ApJ...777...52Z}.
\end{barticle}
\endbibitem

\bibitem[\protect\citeauthoryear{{Zharkova}
  \textit{et~al.}}{2005}]{2005SoPh..228..361Z}
\begin{barticle}
\bauthor{\bsnm{{Zharkova}}, \binits{V.V.}},
\bauthor{\bsnm{{Aboudarham}}, \binits{J.}},
\bauthor{\bsnm{{Zharkov}}, \binits{S.}},
\bauthor{\bsnm{{Ipson}}, \binits{S.S.}},
\bauthor{\bsnm{{Benkhalil}}, \binits{A.K.}},
\bauthor{\bsnm{{Fuller}}, \binits{N.}}:
\byear{2005},
\batitle{{Solar Feature Catalogues In Egso}}.
\bjtitle{\solphys}
\bvolume{228},
\bfpage{361}.
\doiurl{10.1007/s11207-005-5623-0}.
\adsurl{2005SoPh..228..361Z}.
\end{barticle}
\endbibitem

\bibitem[\protect\citeauthoryear{{Zirin}}{1966}]{1966soat.book.....Z}
\begin{bbook}
\bauthor{\bsnm{{Zirin}}, \binits{H.}}:
\byear{1966},
\bbtitle{{The solar atmosphere}},
\bpublisher{Blaisdell Pub. Co.},
\blocation{Waltham, Massachuets}.
\adsurl{1966soat.book.....Z}.
\end{bbook}
\endbibitem

\end{thebibliography}

\end{article}
\end{document}